\documentclass[pra,amsmath,twocolumn,showpacs]{revtex4}
\usepackage{graphicx}
\usepackage{amsmath}
\usepackage{eufrak}
\begin{document}
\def\la{{\langle}}
\def\ra{{\rangle}}
\def\vep{{\varepsilon}}
\newcommand{\beq}{\begin{equation}}
\newcommand{\eeq}{\end{equation}}
\newcommand{\beqa}{\begin{eqnarray}}
\newcommand{\eeqa}{\end{eqnarray}}
\newcommand{\q}{\quad}
\newcommand{\tunn}{\text{tunn}}
\newcommand{\refl}{\text{refl}}
\newcommand{\all}{\text{all}}
\newcommand{\ion}{\text{ion}}
\newcommand{\bound}{\text{bound}}
\newcommand{\free}{\text{free}}
\newcommand{\lc}{\curly{l}}
\newcommand{\A}{\hat{A}}
\newcommand{\B}{\hat{B}}
\newcommand{\s}{\hat{S}}
\newcommand{\x}{x_{cl}}
\newcommand{\si}{\hat{\sigma}}
\newcommand{\Pp}{\hat{\Pi}}
\newcommand{\AC}{{\it AC }}
\newcommand{\La}{{\lambda }}
\newcommand{\psii}{|\psi_I\ra}
\newcommand{\psif}{|\psi_F\ra}
\newcommand{\n}{\\ \nonumber}
\newcommand{\nn}{\q\q\q\q\q\q\q\q\q\\ \nonumber}
\q\q\q\q\q\q\q\q\q
\newcommand{\om}{\omega}
\newcommand{\U}{\hat{U}}
\newcommand{\up}{\hat{U}_{part}}
\newcommand{\mf}{m_f^{\alpha}}
\newcommand{\e}{\epsilon}
\newcommand{\Om}{\Omega}
\newcommand{\Tau}{\mathcal{T}_{SWP}}
\newcommand{\Ttu}{\tau_{in/out}}
\newcommand{\br}{\overline}
\newcommand{\cn}[1]{#1_{\hbox{\scriptsize{con}}}}
\newcommand{\sy}[1]{#1_{\hbox{\scriptsize{sys}}}}
\newcommand{\pd }{Pad\'{e} }
\newcommand{\PAD }{Pad\'{e}\q}
\newcommand{\PP }{\hat{\Pi}}
\newcommand{\get }{\leftarrow}
\newcommand{\f}{\ref }
\newcommand{\T}{\text{T}_\Om}
\newcommand{\Tf}{\text{T}}
\newcommand{{\ttau}}{\overline{\tau_\Om} }
\newcommand{{\tttu}}{\overline{\tau_{[0,d]}} }
\newcommand{\h}{\hat{H}}
\newcommand{\N}{\mathfrak{N} }
\newcommand{\I}{\text{Im } }

\title{ The Salecker-Wigner-Peres clock, Feynman paths, and a tunnelling time that should not exist}
\author {D. Sokolovski$^{a,b}$}
\affiliation{$^a$ Departmento de Qu\'imica-F\'isica, Universidad del Pa\' is Vasco, UPV/EHU, Leioa, Spain}
\affiliation{$^b$ IKERBASQUE, Basque Foundation for Science, Maria Diaz de Haro 3, 48013, Bilbao, Spain}
 
\begin{abstract}
\noindent
The Salecker-Wigner-Peres (SWP) clock is often used to determine the duration  a quantum particle is supposed to spend is a specified region of space $\Om$. By construction, the result is a real positive number, and the method seems to avoid the difficulty of introducing complex time parameters, which arises in the Feynman paths approach. However, it tells little about about the particle's motion. We investigate this matter further, and show that the SWP clock, like any other Larmor clock, correlates the rotation of its angular momentum with the durations, $\tau$, which the Feynman paths spend in $\Om$, thereby destroying interference between different durations.  An inaccurate weakly coupled clock leaves the interference almost intact, and the need to resolve the resulting "which way?"
problem is one of the main difficulties at the centre of the "tunnelling time" controversy. In the absence of a probability distribution for the values of $\tau$, the SWP results are expressed in terms of moduli of the "complex times", given by the weighted sums of the corresponding probability amplitudes. It is shown that over-interpretation of these results, by treating the  SWP times as physical time intervals, leads to paradoxes and should be avoided. We also analyse various settings of the SWP clock, different calibration procedures, and the relation between the SWP results and the quantum dwell time. 
The cases of stationary tunnelling and tunnel ionisation are considered in some detail.
Although our detailed analysis addresses only one particular definition of the duration of a tunnelling process, it also points towards the 
impossibility of uniting various time parameters, which may occur in quantum theory, within the concept of a single "tunnelling time".
\pacs{ 03.65.Xp,73.40.Gk}
\date{\today}
\end{abstract}
\maketitle

\vskip0.5cm
\section{introduction}
Recent progress in attosecond science \cite{UFAST} has returned to prominence  
 the nearly hundred years old \cite{McColl} question "how long does it take for a particle to tunnel?".  There are serious disagreements, e.g.,  between the authors of \cite{ZeroT}, who claimed that "optical tunnelling is instantaneous", and the conclusions of \cite{Teeny1} suggesting that "the electron spends a non-vanishing time under the potential barrier". An overview of the "tunnelling time problem", in its relation to  attosecond physics, can be found, for example, in \cite{LansREV}.
\newline
The tunnelling time problem was extensively investigated in the last decade of the previous century, mostly in the context of tunnelling across stationary potential barriers, closely related to the then-fashionable subject of carrier transport in heterostructures  (for review see \cite{REV1}- \cite{REV2}). The problem also has a more fundamental  aspect.
 An often cited difficulty in defining a tunnelling time is the absence of the corresponding hermitian operator, an the impossibility of performing a standard von Neumann measurement \cite{vN}  in order to determine it.
However, this is not a major obstacle, since the von Neumann procedure can be extended to measuring quantities represented by certain types of functionals on the Feynman paths of the measured system \cite{FUNC1}- \cite{FUNC2}, by making the meter monitor the system over an extended period of time. 
\newline
One time parameter, represented by such a functional, is the net time a quantum particle spends in the specified region of space.
It is intimately related to Larmor precession, and following Buettiker \cite{Buett2} we will refer to it as the {\it traversal time}.
With the functional specified, the problem becomes one in quantum measurement theory. It was studied in some depth in 
\cite{DSB}-\cite{DSbook}. The main conclusion of these studies, which we maintain to date, is as follows.
Traversal time can be measured by an extended von Neumann procedure, and the relevant meter is a variant of a Larmor clock, 
a spin, whose angle of rotation correlates with the duration spent in the magnetic field \cite{QUINT}. 
However, a quantum measurement is significantly more complicated than its classical counterpart,
largely due to the trade-off between its accuracy, and the perturbation the measurement produces. 
Larmor clocks with spins of different sizes, observed in different states and subjected to different magnetic fields, will all produce different results.
These results, although perfectly tractable, lack the universality of the classical traversal time.
To put it differently, analysis of the quantum traversal time problem is worthy from the general point of view, but its result is bound to disappoint a practitioner wishing to know only 
"how many seconds does it take to tunnel, after all?".
\newline
Among many possible versions of the Larmor clock \cite{Buett2}, \cite{Larm1}-\cite{Larm2}, one stands out, and has been the subject of many recent and not so recent studies
\cite{SWP1}-\cite{SWP2}. The Salecker-Wigner-Peres (SWP) clock was first considered as a quantum tool for measuring space-time distances in the general relativity \cite{SWP1}, and was later adopted by Peres \cite{SWP3} for timing events in non-relativistic quantum mechanics.
Specifications of the SWP clock include the choice of its initial and final states, the size of the spin, the strength of the field, and the particular way in which the result of the measurement is calculated.  The resulting time can represented as the average value of  
the "clock time" operator and is, by construction, a real positive number. The SWP result is often taken to be the {definition}
of the time a particle spends in the magnetic field contained in the region of interest.
One reason why the analysis must not stop there is because such a result tells little about the particle's motion.
Timing a classical particle by means of a classical stopwatch, and getting a result of one second, implies that the particle has actually spent one second in the region $\Om$, 
plus all practical consequences one can draw from this information. The implications of measuring one second with a quantum clock remain unclear, until one considers its precise relation to the particle's Feynman paths.
\newline
Like every Larmor clock, the SWP clock modifies the contributions the Feynman paths make to a transition amplitude, depending on the final state in which the clock is found. As one would expect, a nearly classical clock, equipped with a very large spin or angular momentum, 
destroys the interference between the paths spending different durations in $\Om$ almost completely. In this case, 
having found the initial state of the clock rotated by an angle $\phi$, one can be certain that the particle did spend in $\Om$ $\phi/\om_L$ seconds, where $\om_L$ stands for the Larmor frequency \cite{QUINT}. Choosing a weaker field, or a smaller angular momentum, would 
leave certain amount of the interference intact, and reduce the accuracy of the measurement. Even so, 
by varying the accuracy, one can probe certain aspects of the particle's motion. For example, in the case of resonance tunnelling across a double barrier, a measurement of a medium accuracy allows one to identify the long delays associated with 
the exponential decay of the  barrier's metastable state [see Fig.8 of \cite{SBrouard}]. Improving the accuracy, one finds the evidence
of the particle "bouncing" between the potential walls [see Fig.9 of \cite{SBrouard}]. Both the decay and the "bounces" are often 
associated with the particle being trapped in a metastable well. Both can be observed, but not at the same time \cite{SBrouard}. 
One problem with quantum time measurements is that the restrictions on the Feynman paths, imposed by the clock, 
tend to perturb the transition the particle is supposed to make. Thus, if an accurate clock is employed, the particle may either not reach 
its final state at all, or be seen to spend no time in $\Om$ \cite{SBrouard}. Similarly, resonance tunnelling, even in the presence 
of a relatively inaccurate Larmor clock, will not be the same, as without it. Yet when one asks "how long it takes to tunnel?", 
he/she usually means "unperturbed". This is a well known difficulty in quantum mechanics, where "to know" often implies "to disturb". 
\newline
A natural way to avoid the unwelcome perturbation is to reduce the coupling between the clock and the system, and try to interpret whatever information can be gained in this manner. The purpose of this paper is to analyse the results obtained by an SWP clock in the limit $\om_L \to 0$, and relate them to the time parameters describing the motion of a quantum particle, involved in a  transition between known initial and final states. This brings the discussion into the realm of the inaccurate, "indirect" \cite{QUINT}, or "weak" \cite{LANprl} measurements of the traversal time. In the "weak" regime, we can expect a weak SWP clock to make a rather poor job of destroying interference between different values of the traversal time. We will also need to heed D. Bohm's warning \cite{Bohm} that "if the interference were not destroyed", "the quantum theory could be shown to lead to absurd results", and see what it means for the quest to find "the tunnelling time". 
\newline
The rest of the paper is organised as follows.
\newline
In Section II we discuss various time parameters describing the motion of  a classical particle.
\newline
Section III lists some of the quantum time parameters which are not discussed in this paper.
\newline
In Sect. IV we define the quantum traversal time, and its amplitude distribution, for a particle pre- and post-selected in the known initial and final states.
\newline
In Sect. V we introduce the "complex times", which are likely to arise in any weakly perturbing measurement scheme.
\newline  
In Sect. VI we cast  the complex times into a more familiar operator form. 
\newline
In Sect. VII we describe the family of Larmor clocks, and their relation to the amplitude distribution of the quantum traversal time.
\newline
In Sect. VIII we introduce the SWP clock as a particular member of the family.
\newline
In Sect. IX we reduce the coupling, and show that the time measured by a weakly coupled SWP clock is naturally 
expressed in terms the moduli of the complex time of Sect. V.
\newline
In Sect. X we study the calibration procedure proposed in \cite{Leav1}, and demonstrate that it can lead to "absurd results" predicted by Bohm.   
\newline
In Sect. XI we try to make sense of these "absurd results", and establish a connection between the complex times and
the weak values of quantum measurement theory.
\newline
In Sect. XII we revisit the dwell time and show it to be a particular case of the "complex times" of Sect. V.
\newline
In Sect. XIII we ask whether the Peres'  clock would measure the dwell time, and find that it would not.
\newline
In Sect. XIV we apply our general analysis to tunnelling across a stationary potential barrier.
\newline
In Sect. XV we apply the analysis to a simple model of tunnel ionisation.
\newline
Section XVI contains our conclusions.
\section{Which classical time?}
We start by reiterating the three questions which, in our opinion, one might want to answer before performing a quantum measurement. These are:

(i) What is being measured?

(ii) By what means is it being measured?

(iii) To what accuracy is it being measured?
\newline
The first question arises already in classical mechanics, when we discuss the time parameters describing the presence of a particle, moving along a trajectory $\x(t)$, in a specified region of space, $\Om$. One obvious choice is the net duration the particle spends in $\Om$. It is given by the integral \cite{DSB}
\begin{eqnarray}\label{a1}
\tau_\Om[\x(t)]=\int_{t_1}^{t_2}\Theta_{\Om}(\x(t))dt,
\end{eqnarray}
where $\Theta_{\Om}(z)$ has the value of one for a $z$ inside $\Omega$, and zero otherwise. 
Another choice would be, for example, the time interval between the moments  $t_{in}$, when the particle enters $\Om$ for the first time,
and  $t_{out}$, when it leaves it for the last time, 
\begin{eqnarray}\label{a2}
\Ttu[\x(t)]=t_{out}[\x(t)]-t_{in}[\x(t)].
\end{eqnarray}
The choice depends on the question we want to ask. If the particle has a tendency to change colour from white to black 
proportionally to the net duration spent in $\Om$, to predict the shade of grey acquired we need $\tau_\Omega [\x(t)]$.
 If, on the other hand the temperature in $\Om$  changes with a frequency $\omega$, and we need the particle 
to experience no change, the required condition would be $\omega\Ttu[\x(t)]<<1$, and not  $\omega\tau_\Om[\x(t)]<<1$. 
In general, these two parameters are different.
\begin{eqnarray}\label{a3}
\tau_\Om[\x(t)] \ne \Ttu[\x(t)].
\end{eqnarray}
Even classically, different time parameters require different measurement procedures. To measure $\tau_\Om[\x(t)]$, we can equip 
the particle with a magnetic moment which precesses in the magnetic field introduced in $\Om$. Dividing the final angle of precession by the Larmor frequency, $\omega_L$, we obtain the value of $\tau_\Om[\x(t)]$. It appears that no similar procedure exists for $\Ttu[\x(t)]$, or even for $t_{in}[x(t)]$ in Eq.(\ref{a2}) \cite{FUNC4}. The difficulty is in stopping the clock after the {\it first} entry in $\Om$, and preventing it from running again should the path leave the region and then re-enter. In classical mechanics we can simply plot the particle's trajectory, and determine $t_{in}[\x(t)]$ from the graph. In the quantum case,  there is no trajectory to draw, and the absence of a meter is a serious problem \cite{FUNC4}.
\newline
The accuracy of a measurement is of no great importance in the classical case, where a meter (a clock) can monitor a particle 
with any precision, without altering its trajectory $\x(t)$. It plays a much more important role in the quantum case, where there is a tradeoff between the accuracy of the measurement, and the perturbation the meter exerts on the particle's motion. 
\newline 
Throughout the rest of the paper we will try to answer the following question: {\it What is the total amount of time a quantum particle starting in a known state $\psii$ at $t=t_1$, and then observed on a state $\psif$ at $t=t_2$, had spent in a specified region $\Om$ 
between $t_1$and $t_2$? }. To measure it we will employ a highly inaccurate Salecker-Wigner-Peres clock, 
specifically designed to perturb the studied quantum transition as little as possible.
The experiment we have in mind is like this. A particle is prepared in $\psii$, coupled to an SWP clock, and then detected in $\psif$.
If the detection is successful, we "read" the clock in some manner, record the result, and draw conclusions about the duration spent in $\Om$. Although we consider one-dimensional scattering, most of our results can be extended to two or three dimensions.
 \section{Other quantum times  beyond the scope of this paper} 
 In quantum mechanics there are many different ways to introduce quantities measured in units of time.
 Before proceeding with our main task, we briefly discuss some of the time parameters, which describe a scattering (tunnelling) process and are {\it not } a subject of of this paper. 
 \newline
The simplest way to probe the tunnelling delay is to prepare a particle in a wave packet state on one side of the barrier, choose a location $x$ on its other side, and evaluate the probability $P(x,t)=|\psi(x,t)|^2$.  
Using $P(x,t)$ as a probability distribution, one can construct the 
real non-negative mean time \cite{Japha}, also known as the "time of presence" \cite{Muga1}
\begin{eqnarray}\label{p1}
\la t(x)\ra = \int t P(t,x)dt/\int P(x,t)dt.
\end{eqnarray} 
This mean time can be measured by performing $N>>1$ trials, each time checking whether the particle is between $x$ and $x+dx$
at a time $t$. If in $N_1$ cases the particle is found there, the ratio $N_1/Ndx$ would yield an approximate value of $P(x,t)$.
Repeating the checks at various times, allows one to reconstruct $P(x,t)$ and, with it, $\la t(x)\ra$.
This is, however, different from what we intend to do here, as explained in Sect. 18 of \cite{DSann}.
\newline
A slightly different method was recently proposed by Pollack in Refs. \cite{Poll1}, \cite{Poll2}. There
the particle is prepared in a thermal mixed state
\begin{eqnarray}\label{p1a}
\hat{\rho}_I= \exp(-\beta \h/2)|x_0\ra\la x_0|\exp(-\beta \h/2),
\end{eqnarray}
where $x_0$ is some initial location, $\h$ is the Hamiltonian, and $\beta$ is the inverse temperature. The state is evolved until some $t$, $\hat{\rho}(t)=\exp(i\h t)\hat{\rho}_I\exp(-i\h t)$, and the probability to find the particle at a location $x$ on the other side of the barrier, $P(x,t)=tr\{|x\ra\la x| \hat{\rho}(t)\}$, is inserted into Eq.(\ref{p1}) for the mean transit time. The mean time can then be measured as discussed above. 
This is also not what we wish to discuss below, if only because here we are not interested in systems in thermal equilibrium. 
\newline
Finally, the authors of \cite{Teeny1} proposed using the probability current evaluated at two locations  on the opposite sides of the barrier, $x_1$ and $x_2$, and define the mean transit time as the difference between the moments the out- and in-going probability currents at $x_2$ and $x_1$ reach their maxima.
Measuring, albeit indirectly, this time would require a different experiment, e.g., the one in which the presence of the particle is
checked at all times to the right of $x_1$, and then at $x_2$, the evaluated probabilities are differentiated with respect to time to yield the currents, and the maxima of the two curves are identified. This procedure is not our subject either.

The list of possible quantum time parameters can be extended, and new times will, undoubtably, be proposed in future studies.
It is not our intention to compare relative merits or defects of the approaches discussed in this Section. (Except, perhaps, 
citing some of well known problems with defining quantum arrival times \cite{Muga1}, or relying on the probability current in order to determine times, or time intervals \cite{Muga2}). Rather, we note that measurements of different quantum times require different experimental procedures, and should not be expected to give the same result. To some extent this is true already in classical mechanics, 
as was pointed out in the previous Section. Thus, A may propose, and perform, an experiment in which a time parameter associated with a tunnelling transition vanishes, and claim tunnelling to be an "infinitely fast" process. B can do something different, obtain a non-zero answer, and state "that tunnelling does take time after all". The argument between  A and B will never have a meaningful resolution, since both claims rely on the assumption that there is a single time tunnelling "takes", and there is overwhelming evidence that this assumption is false. 
In this paper, to add to this evidence, we consider a particular classical time (\ref{a1}), and see what will happen if it is generalised to the full quantum case.
 \section{Traversal time for quantum motion}  
A classical particle of a mass $\mu$ in a potential $V(x,t)$ goes from some initial position $x_I$ at $t=t_1$ to a final position $x_F$ at $t=t_2$ along a smooth continuous trajectory $x_{cl}(t)$. There is a single value of the duration spent in $\Om$, and it is given by the functional  $\tau_\Om[x(t)]$ in Eq.(\ref{a1}).
\newline
The quantum case is more complex. A quantum particle can make a transition from an initial state $|\psi_I\ra$ at $t=t_1$ to a final state $|\psi_F\ra$ at $t=t_2$. To proceed, we need to choose a representation. Since we are interested in a spacial region $\Om$, 
the coordinate representation is the appropriate one.  Now a point particle can be thought of as being at some location $x(t)$ at any time $t_1\le t\le t_2$, 
and a possible scenario for reaching $\psi_F$ from $\psi_I$ is by following a Feynman path $x(t)$, which is continuous, but not smooth \cite{Feyn}. The path is virtual, and is equipped only with a probability amplitude (we use $\hbar=1$)
\begin{eqnarray}\label{b1}
A([x(t)],\psi_I,\psi_F)=\la \psi_F|x_F\ra \exp\{iS[x(t)]\}\la x_I |\psi_I\ra,
\end{eqnarray}
where 
 $S[x(t)]=\int_{t_1}^{t_2}[\dot{x}^2/2\mu-V(x,t)]dt$ is the classical action.
The full transition amplitude to reach $\psif$ from $\psii$ is given by the Feynman path integral \cite{Feyn}, which we symbolically write  as
\begin{eqnarray}\label{b2}
A(\psi_F, \psi_I, t_2,t_1)= \sum_{paths}A([x(t)],\psi_I,\psi_F).
\end{eqnarray} 
Note that the set of Feynman paths in Eq.(\ref{b2}) is always the same. What changes, with the change of the potential in which a particle moves, are the path amplitudes $A([x(t)],\psi_I,\psi_F)$. The classical dynamics emerges from Eq.(\ref{b2}) when the contribution to 
the path integral comes from the vicinity of the path $x_{cl}(t)$ on which $S[x(t)]$ is stationary \cite{Feyn}.
\newline
 What can be said about the duration a particle spends in $\Om$ is dictated by the basic rules of quantum mechanics.
 The functional $\tau_\Om[x(t)]$ can be evaluated for each of the Feynman paths. The paths can be combined and recombined into new pathways, just as the superposition principle allows us to recombine vectors in Hilbert space into  a new vector \cite{FUNC2}. Combining together all the paths which share the same value $\tau$ of  
 $\tau_\Om[x(t)]$, we create a new virtual pathway, {\it for reaching} $\psif$ {\it from} $\psii$, {\it and spending} $\tau$ {\it seconds in} $\Om$ {\it along the way}, and sacrifice to interference all other information contained in the individual Feynman paths.
The amplitude for the new pathway is 
\begin{eqnarray}\label{b3}
A(\psi_F, \psi_I, t_2,t_1|\tau)= \n
\sum_{paths}A([x(t)],\psi_I,\psi_F)\delta(\tau_\Om[x(t)]-\tau),
\end{eqnarray} 
where $\delta(z)$ is the Dirac delta. Integrating Eq.(\ref{b3}) over all possible $\tau$'s restores the full transition amplitude
$A(\psi_F, \psi_I, t_2,t_1)$ in Eq.(\ref{b2}).In addition, we have
 \begin{eqnarray}\label{b3a}
A(\psi_F, \psi_I, t_2,t_1|\tau)\equiv 0, \q \text{for} \q \tau<0 \q \n
 \text{and} \q \tau > t_2-t_1,
\end{eqnarray} 
since non-relativistic Feynman paths may not spend in $\Om$ a duration which is either negative, or exceeds 
the total duration of motion.
\newline
The situation is a standard one in quantum mechanics. For given initial and final states of the particle, we have not one, but infinitely many values of the traversal time $\tau$. 
To each value we can ascribe a probability amplitude, but not the probability itself. 
This is not different from what happens in Young's two-slit experiment \cite{QUINT}. 
The expectation that there must, after all, be a single traversal time associated with a quantum transition, is as good, or as bad, as the assumption that each electron must have 
actually gone through one slit or another. According to Feynman \cite{FeynLaw}, the latter assumption should be abandoned, and
the rule for adding amplitudes must be accepted as the basic axiom of quantum theory instead.
Throughout the rest of the paper, we will maintain this point of view, despite possible objections from the proponents of the Bohmian version of quantum theory  \cite{Leav3}, \cite{Leav4}, \cite{Holl}.
\newline
Thus, our virtual pathways, labelled by the value of $\tau$, interfere just like individual Feynman paths they comprise, 
and should together be considered a single indivisible pathway connecting $\psii$ and $\psif$ \cite{FUNC2}.
Interference between them can be destroyed by an accurate meter \cite{QUINT}, registering the actual value of $\tau$ each time the transition is observed, but the probability $P_{acc}(\psi_F, \psi_I, t_2,t_1)$ to reach the final state, while being observed,  will change, 
\begin{eqnarray}\label{b4}
P_{acc}(\psi_F, \psi_I, t_2,t_1) \equiv \int_{0}^{t_2-t_1} d\tau |A(\psi_F,\psi_I, t_2,t_1|\tau)|^2 \ne \q\q\n
|\int_{0}^{t_2-t_1} d\tau A(\psi_F,\psi_I, t_2,t_1|\tau)|^2=|\sum_{paths}A([x(t)],\psi_I,\psi_F)|^2.
\end{eqnarray} 
This simple discussion should help to establish the status of the time parameter represented by the functional (\ref{a1}) within the standard quantum theory. We note the similarity between the quantum traversal time problem and the Young's double-slit experiment.
If we were able to construct a unique traversal time, or even a probability distribution for such times, we could, in principle, also determine the slit through which the electron has passed, with the interference pattern on the screen intact. 
According to Feynman \cite{FeynLaw}, the latter is an impossible task.
\section{The complex times} 
Even before going into the details of a particular measurement, 
we can guess what would happen if the meter's 
 interaction with the particle has been deliberately made small (weak), in order to preserve the interference, and minimise the perturbation produced on the particle's motion.
There is a fashionable view that the result would be a "weak value", a new type of quantum variable, capable of providing a new insight into physical reality. (For a recent review see \cite{WEAK11}, the term "weak measurement elements of reality" was coined in \cite{WEAK12}).
\newline
 Recently we argued against over-interpretation of the "weak values", and offered a more prosaic explanation  \cite{FUNC2}, \cite{PLA2016}.
In the absence of probabilities, any weakly  perturbing  scheme is bound to give a result, expressed in terms of the probability amplitudes. A scheme set up to weakly measure a quantity would typically yield a {\it real} result expressed, in one way or another,  in terms of the complex valued {\it sums of the 
corresponding amplitudes, weighed by the values of the measured quantity}
 and, occasionally, the amplitudes themselves  \cite{FUNC2}, \cite{PLA2016}. 
Far from representing a new type of "reality", these results only give us this limited information about the particular set of virtual pathways connecting the initial and final states. 
They would, for example, shed no new light on the mechanism of the two-slit experiment, mentioned in the previous Section, beyond what is known from the textbooks.
\newline
In the case of the traversal time (\ref{a1})  one such weighted sum is the "complex time" introduced in \cite{DSB}, 
\begin{eqnarray}\label{b5}
\overline{\tau_{\Om}}(\psi_I,\psi_F)=\q\q\q\q\q\q\q\q\q\n
\sum_{paths}\tau_\Om[x(t)]A([x(t)],\psi_I,\psi_F)/\sum_{paths}A([x(t)],\psi_I,\psi_F) \q\q \n
=\int_{0}^{t_2-t_1} d\tau \tau A(\psi_F,\psi_I, t_2,t_1|\tau)/A(\psi_F, \psi_I, t_2,t_1). \q\q
\end{eqnarray}
The quantities of these type, first introduced by Feynman \cite{Feyn} as  "transition elements of functionals", 
reduce to the weak values of \cite{WEAK11}, if the functional in question is the instantaneous value of a variable 
$A(t_0)$ at a time $t_1<t_0<t_2$.
\newline
The quantity in Eq.(\ref{b5}) was often dismissed as a candidate for the duration quantum particle performing a transition (e.g., tunnelling transmission 
across a potential barrier) spends in a specified region of space, on account of it being complex valued. For example, in Ref.\cite{REV1} we read: "...common sense dictates that to the question of the duration of a tunneling
process, the answer, if it exists at all, must be a real
time". The key words here are "if it exists", and in the previous Section we explained in what sense the answer should not exist.
As a consequence, only complex valued combinations of transition amplitudes similar to (\ref{b5}) will be found in the analysis of a non-perturbing weakly coupled meter. It is true that the result of a physical measurement must be real, but there is no contradiction. Different setups, employed to weakly measure the quantum traversal time (\ref{a1}), may yield $\text{Re} \ttau_{}$, $\text{Im} \ttau_{}$ or $|\ttau_{}|$, as demonstrated in the table 
in \cite{DSwp}. 
\newline
Finally, we can define expressions similar to (\ref{b5}) for higher powers of the functional $\tau[x(t)]$,
\begin{eqnarray}\label{xy5}
\overline{\tau^n_\Om}(\psi_I,\psi_F,t_2,t_1)
\equiv \frac{\int_{0}^{t_2-t_1} d\tau \tau^n A(\psi_F,\psi_I, t_2,t_1|\tau)}{A(\psi_F, \psi_I, t_2,t_1)}, \q 
\end{eqnarray}
where $n=2,3....$. We will require some of these quantities in what follows.
In the following Sections we illustrate what has been said so far, using the example of a weakly coupled SWP clock. 
\section{Complex times in operator notations}
It may be convenient to formulate the problem in terms of partial evolution operators acting in the particle's Hilbert space. 
Let $\up[x(t)] =|x_F\ra \exp\{iS[x(t)]\} \la x_I|$ be an operator which evolves the (part)icle along a single Feynman path $x(t)$, which starts in $x_I$ at $t_1$,and ends in $x_F$ at $t_2$. Summing over all paths which spend exactly $\tau$ seconds in $\Om$, (summation over $x_I$ and $x_F$ included), we obtain an evolution operator {\it conditioned} by the requirement that the particle spend exactly $\tau$ seconds in the region of interest, 
\begin{eqnarray}\label{xy1}
\up(t_2,t_1|\tau)=\sum_{paths}\up[x(t)]\delta(\tau[x(t)]-\tau).
\end{eqnarray}
Summing Eq.(\ref{xy1}) over all $\tau$'s restores the full evolution operator, $\up(t_2,t_1)$,
\begin{eqnarray}\label{xy2}
\int_{0}^{t_2-t_1}\up(t_2,t_1|\tau)d\tau =\sum_{paths}\up[x(t)]\n
=\up(t_2,t_1).\q\q
\end{eqnarray}
Next we introduce an operator $\up(t_2,t_1|\La)$ as a Fourier transform of $\up(t_2,t_1|\tau)$,
\begin{eqnarray}\label{xy3}
\U_{part}(t_2,t_1|\tau)=
(2\pi)^{-1/2}\int \exp(i\lambda \tau) \up(t_2,t_1|\La)d\La,\q\q 
\end{eqnarray}
and, with it,  an operator family, 
\begin{eqnarray}\label{xy4}
\U_{part}^{(n)}(t_2,t_1)=\int_{0}^{t_2-t_1} \tau^n\U_{part}(t_2,t_1|\tau) d\tau \n
=(i)^n \partial^n_\La \U_{part}(t_2,t_1|\La)|_{\La=0},\q n=0,1,2...,
\end{eqnarray}
where $\U_{part}^{(0)}(t_2,t_1)=\U_{part}^{}(t_2,t_1)$.
\newline Now the complex "averages" in Eqs. (\ref{b5}) and (\ref{xy5}), 
can also be written as 
\begin{eqnarray}\label{xy6}
\overline{\tau^n_\Om}(\psi_F, \psi_I, t_2,t_1)=
\frac{\la \psi_F|\psi^{(n)}\ra}{\la \psi_F|\psi^{(0)}\ra},\q 
\end{eqnarray}
where 
\begin{eqnarray}\label{xy7}
|\psi^{(n)}(t_2,t_1,\psi_I)\ra \equiv \U^{(n)}_{part}(t_2,t_1)|\psi_I\ra,  
\end{eqnarray}
and $\overline{\tau_{\Om}}(\psi_I,\psi_F)=\overline{\tau^{(1)}_{\Om}}(\psi_I,\psi_F)$.
\newline
The usefulness of this approach becomes more evident as we realise that $\up(t_2,t_1|\La)$ coincides with the evolution operator 
for a particle moving in the original potential $V(x,t)$ plus an additional potential which equals $\La$ inside $\Om$, and 
vanishes outside it, $\up(t_2,t_1|\La)=\exp[-i\int_{t_1}^{t_2}\h_{part}(t,\La)dt]$, where the particle's Hamiltonian, $\h_{part}(t,\La)$, is given by
\begin{eqnarray}\label{xy8}
\h_{part}(t,\La) \equiv -\partial_x^2/2\mu + V(x,t)+\lambda \Theta_\Om(x).
\end{eqnarray}
This result follows by noting that
if one writes $\delta(\tau[x(t)]-\tau)$ as $(2\pi)^{-1}\int \exp\{i\La (\tau-\tau[x(t)])\}d\lambda$ and inserts it in (\ref{b3}), the action 
$S[x(t)]$ in Eq.(\ref{b1}) is modified by the term $-\La \int_{t_1}^{t_2}\Theta_\Om(x(t))dt$, which corresponds to adding an extra potential 
$\La \Theta_\Om(x)$. The operators $\U_{part}^{(n)}(t_2,t_1)$ can now be evaluated by expanding $\up(t_2,t_1|\La)$ in powers
of $\La$ with the help of the perturbation theory \cite{Feyn}. For example, for $\U_{part}^{(1)}(t_2,t_1)$ we have
\begin{eqnarray}\label{xy9}
\U_{part}^{(1)}(t_2,t_1) =\q\q\q\q\q\q\n
\int_{t_1}^{t_2}dt'\int_\Om dx' \up(t_2,t')|x'\ra\la x'|\up(t',t_1),\q\q
\end{eqnarray}
so that
\begin{eqnarray}\label{xy10}
\overline{\tau_\Om}(\psi_I,\psi_F)=
\frac{\int_{t_1}^{t_2}dt'\int_\Om dx' \psi^*_F(t',x')\psi_I(t',x')}{\la \psi_F|\up(t_2,t_1)|\psi_I\ra}, 
\end{eqnarray}
where $\psi_I(t',x')\equiv \la x'|\up(t',t_1)|\psi_I\ra$ and $\psi_F(t',x')\equiv \la x'|\up^{\dagger}(t_2,t')|\psi_F\ra$.
With the help of Eq.(\ref{xy6}) 
 is easy to prove an identity
\begin{eqnarray}\label{A10}
\la \psi^{(m)}| \psi^{(n)}\ra = \overline{\tau^n_\Om}(\psi^{(m)},\psi_I) \overline{\tau^{m}_\Om}^*(\psi^{(0)},\psi_I),
\end{eqnarray} 
which we will use in what follows.
\newline
Finally, the Fourier transform (\ref{xy3}), relating $\up(t_2,t_1|\tau)$ to $\up(t_2,t_1|\La)$, suggests that $\tau$ and $\La$ are,
in some sense, "conjugate variables". They must satisfy an uncertainty relation \cite{SBrouard}, so that a narrow amplitude distribution of $\tau$ would imply a broad range of $\La$'s, and vice versa. 
Thus, in order to know the duration $\tau$ spent in $\Om$, one must make the potential in $\Om$ uncertain. Conversely, if the potential
is sharply defined, $\tau$ cannot, in general, be known exactly. For example, the choice $\up(t_2,t_1|\tau)=\up(t_2,t_1)\delta(\tau - \tau_0)$ makes  in Eq.(\ref{xy3}) the effective range of integration over $\La$ infinite. Therefore an evolution for which $\tau$ is known exactly, must be represented as a sum of evolutions for all possible potentials $\La\Theta_\Om(x)$ added to $V(x,t)$. It also means that in order 
to measure $\tau$, a meter would need to introduce at least some uncertainty in the potential in the region of interest.
This can be done by equipping the particle with a magnetic moment, proportional to its spin, or angular momentum, 
so that each component of the spin would experience a different potential inside $\Om$, where a constant magnetic field is introduced. There are various ways for preparing the spin degree of freedom, and we will discuss them next.
\section{The family of Larmor clocks}
Any spin-rotating (Larmor) quantum clock relies on the fact that a magnetic moment, proportional to a spin of a size $j$,  undergoes in a magnetic field Larmor precession with an angular frequency $\om_L$. 
Let the field be directed along the $z$-axis, and $|m\ra$ denote the state in which the projection of the spin on the axis is $m$, so that the spin's Hamiltonian is given by $\h_{spin}=\om_L \hat{j}_z$, with $\hat{j}_z|m\ra=m|m\ra$.
Then, after a time $t$,  an arbitrary $(2j+1)$-component initial spin state $|\gamma^I\ra =\sum_{m=-j}^j\gamma_m^I |m\ra$, will end up rotated around the $z$-axis by an angle $\om_Lt$, $|\gamma^I\ra \to |\gamma (t)\ra=\sum_{m=-j}^j\gamma_m^I \exp(-im\om_Lt)|m\ra$.
\newline
If the magnetic field is introduced only in the region $\Om$, and the spin is travelling with a classical particle which follows a trajectory $\x(t)$ for $t_1 \le t \le t_2$, the state rotates only while the particle remains inside $\Om$. The final angle of rotation is 
$\om_L\tau_\Om[\x(t)]$, and we have
 \begin{eqnarray}\label{c1}
|\gamma (t_2)\ra=\exp\{-i\om_L\tau_\Om[\x(t)]\hat{j}_z\}|\gamma^I\ra.
\end{eqnarray}
Generalisation to the case where the particle is quantum, rather than classical, is now straightforward. A transition between 
$|\psi_I\ra$ and $|\psi_F\ra$ involves a range of durations, each occurring with the probability amplitude (\ref{b3}). 
Hence, the final state of the spin is a superposition of all possible rotations weighted by the corresponding amplitudes, 
 \begin{eqnarray}\label{c2}
|\gamma (t_2)\ra=\int_{0} ^{t_2-t_1} A(\psi_F, \psi_I, t_2,t_1|\tau) \n
\times \exp\{-i\om_L\tau\hat{j}_z\}|\gamma^I\ra d\tau.
\end{eqnarray}
The amplitude to find the spin in some state $|\beta\ra =\sum_{m=-j}^j\beta_m |m\ra$ takes a particularly simple form \cite{QUINT},
\begin{eqnarray}\label{c3}
\la \beta |\gamma (t_2)\ra=\int_{0}^{t_2-t_1}G(\om_L\tau|j,\beta,\gamma^I)
A(\psi_F, \psi_I, t_2,t_1|\tau)d\tau,\q\q 
\end{eqnarray}
where
\begin{eqnarray}\label{c4}
G(\om_L\tau|j,\beta,\gamma^I)\equiv \la \beta|\exp(-i\om_L\tau\hat{j}_z|\gamma^I\ra\n
=\sum_{m=-j}^j \beta_m^*\gamma^I_m\exp(-im\om_L\tau).
\end{eqnarray}
Choosing orthonormal bases $|\beta^k\ra$, $k=0,1,..2j$, and $|N\ra$ to describe the spin and the particle \cite{FOOTbox}, respectively, 
we easily reconstruct the state into which the system, initially described by the  product $|\psi_I\ra|\gamma^I\ra$,  evolves
by $t=t_2$, 
\begin{eqnarray}\label{c5}
|\Psi(t_2)\ra=\sum_N \sum_k \int_{0}^{t-t_1}
G(\om_L\tau|j,\beta^k,\gamma^I)
\n\times 
A(N, \psi_I, t_2,t_1|\tau)d\tau|\beta^k\ra|N\ra\q\q 
\end{eqnarray}
Expanding $G$ in a Taylor series around $\om_L=0$, and using the operators of the previous Section, we can rewrite 
this as
\begin{eqnarray}\label{c5b}
\la \beta^k|\Psi(t_2)\ra=
\sum_{n=0}^\infty\frac{(-i\om_L)^n}{n!}\la \beta^k|\hat{j}_z^n|\gamma^ I\ra 
\up^{(n)}(t_2,t_1)|\psi_I\ra.\q\q 
\end{eqnarray} 
We note that in the classical limit, where there is a single classical trajectory $x_{cl}(t)$, and a single classical duration 
$\tau_{cl}=\tau_\Om[x_{cl}]$, a Larmor clock ceases to affect the particle's motion, and is driven by it. Indeed, with $A(N, \psi_I, t_2,t_1|\tau)=A(N, \psi_I, t_2,t_1)\delta(\tau-\tau_{cl})$, from Eq.(\ref{c5}) we have
\begin{eqnarray}\label{c5a}
|\Psi(t_2)\ra=
\exp(-i\om_L \tau_{cl}\hat{j}_z)|\gamma^I\ra \up(t_2,t_1)|\psi_I\ra 
\end{eqnarray}
Transition to the classical limit is discussed, for example, in \cite{SBrouard}. In this limit $A(N, \psi_I, t_2,t_1|\tau)$ becomes highly oscillatory everywhere except in a small vicinity of $\tau=\tau_\Om[\x(t)]\equiv \tau_{cl}$, making $A(N, \psi_I, t,t_1|\tau)$ tend to $A(N, \psi_I, t_2,t_1)\delta(\tau-\tau_{cl})$. This is the only case in which a uniquely defined traversal time can be ascribed to a quantum transition.
\newline
Thus,we have a family of Larmor clocks, each defined by a particular choice of  $\om_L$, $j$, $|\beta^k\ra$, and $|\gamma^I\ra$. 
\section{The Salecker-Wigner-Peres clock} 
A particular choice due to Salecker and Wigner \cite{SWP1}, and also to Peres \cite{SWP3}, defines the Salecker-Wigner-Peres  clock.
After the measurement, the spin is to be observed in one of the orthogonal states, obtained from its initial state,
\begin{eqnarray}\label{d1}
|\gamma^I\ra = |\beta^0\ra= (2j+1)^{-1/2}\sum_{m=-j}^j|m\ra,
\end{eqnarray}
by rotation through one of the angles $\phi_k=2\pi k/(2j+1)$, $k=0,1,...,2j$,
\begin{eqnarray}\label{d2}
|\beta^k\ra = \exp(-i\hat{j}_z\phi_k)|\beta^0\ra=\sum_{m=-j}^j\frac{\exp(-im\phi_k)}{(2j+1)^{-1/2}}|m\ra.\q \n
\end{eqnarray}
The function $G$ in Eq.(\ref{c4}) is, therefore, given by \cite{QUINT}, \cite{SBrouard}
\begin{eqnarray}\label{d2a}
G_{SWP}(\om_L\tau|j,\beta^k,\beta^0)=(2j+1)^{-1}\times\n
\frac{\sin[(2j+1)(\phi_k-\om_L\tau)/2]}{\sin[(\phi_k-\om_L\tau)/2]}.
\end{eqnarray}
A particle may be observed (post-selected) in a particular sub-space $\N$ of its Hilbert space, 
specified by a projector 
\begin{eqnarray}\label{d3}
\PP(\N) = \sum_{N \in \N}|N\ra\la N|.
\end{eqnarray}
The options range from detecting the particle in a single final state $|N_0\ra$ of the chosen basis, $\PP(\N) = |N_0\ra\la N_0|$, 
to being completely ignorant of its final state, thus choosing $\PP(\N) =1$. In all cases, the measured time is {\it defined} as the average 
of the times corresponding to rotations by the angles $\phi_k$, $\tau_k=\phi_k/\om_L$, weighed by the probabilities, $P(k,\N)$, 
for finding the spin in the rotated state,
\begin{eqnarray}\label{d4}
\T(\N,\psi_I)\equiv \sum_{k=0}^{2j}\tau_k P(k,\N)= \sum_{k=1}^{2j}\frac{2\pi k}{(2j+1)\om_L} P(k,\N),\q\q\q
\end{eqnarray}
where 
\begin{eqnarray}\label{d5}
P(k,\N)=\frac{\la \Psi(t_2)|\beta^k\ra\la \beta^k|\PP(\N)|\Psi(t_2)\ra}{\la \Psi(t_2)|\PP(\N)|\Psi(t_2)\ra}, 
\end{eqnarray}
with $|\Psi(t_2)\ra$ given by Eq.(\ref{c5}).
\newline
The rational behind Eqs.(\ref{d4}) and (\ref{d5}) is simple. After a time $t$, the hand of a classical clock rotates by 
a well defined angle, and points at the hour. The final position of a quantum state, which replaces the classical hand, appears 
to be distributed, pointing at different "hours" with different probabilities. Equation (\ref{d4}) represents the "mean time" 
measured in this way, and associated with the passage of the particle from the state $|\psi_I\ra$ to anywhere in the part of its Hilbert space denoted as $\N$.
\newline
There are at least three remarks to be made. Firstly, finding the clock in a state $|\beta^k\ra$, rotated by $\phi_k$, by no means guarantees that the particle has indeed spent a duration $\phi_k/\om_L$ in $\Om$. Unless $G$ in Eq.(\ref{d2a}) is proportional to 
$\delta(\tau-\phi_l/\om_L)$, various durations $\tau$ continue to interfere, 
and the precise time the particle spends in $\Om$ remains, in general,  indeterminate.
\newline
Secondly, if we are not interested in the final state of the particle, $\PP(\N)=1$, $\T(all,\psi_I)$ can be written as an expectation value of an operator $\hat{T}_\Om=\sum_{k=0}^{2j}|\beta^k\ra \tau_k\la \beta^k|$, 
\begin{eqnarray}\label{d4a}
\T(\all,\psi_I)=\la \Psi(t_2)|\hat{T}_\Om|\Psi(t_2)\ra.\q
\end{eqnarray}
It is tempting to conclude that $\hat{T}_\Om$ represents the "traversal time operator", and that with it the "time problem" has been brought into the framework of standard quantum mechanics. This is not quite so, since in quantum measurements the measured operator 
acts on the variables of the studied system, whereas $\hat{T}_\Om$ acts on the variable of the clock.
\newline
Finally, being coupled to the clock perturbs the particle's motion, and whatever information is obtained, no longer refers
to the particle "on its own". The obvious way out of this last difficulty is to try to reduce the coupling as much as it is possible.
In the next Section we will show that this would inevitably lead to "complex times", whose appearance we have already anticipated in Sect. V. 
\section{A non-perturbing (weak) SWP clock}
We still need to specify the values of $\omega_L$ and $j$, which determine the accuracy of the measurement. 
In \cite{QUINT} it was shown that if $j\to \infty$ while $\om_L$ is kept finite, the function $G(\om_L \tau|j,\beta^k,\beta^0)$ in Eq.(\ref{c4}) becomes proportional to $\delta(\tau-\tau_k)$, so that the spin can be found in $|\beta_k\ra$ if, and only if, the particle has actually spent in $\Om$ a duration $\tau$. This is a very accurate measurement, and $P(k,\N)$ in Eq.(\ref{d4}) becomes also the {\it probability}
with which a duration $\tau$ would occur. But, as Eq.(\ref{b4}) demonstrates, by gaining in accuracy we destroyed the original transition, and will have learnt very little about the duration spent in $\Om$ with the interference intact. 
\newline
To pin down this elusive duration we may try to keep $j$ finite, while making the magnetic field, and with it $\om_L$, very small. 
This will reduce the perturbation which affects the particle's motion. Indeed, 
and as $\om_L \to 0$, from Eq.(\ref{c5b})
we have
\begin{eqnarray}\label{c3a}
\la \beta^k |\Psi(t_2)\ra\approx\la \beta^k|\beta^0\ra \up(t_2,t_1)\psii\n
=\delta_{k0}\up(t_2,t_1)\psii. \q
\end{eqnarray}
The particle moves unimpeded, the spin does not rotate, and the clock provides no information at all.
\newline
We need to find the first correction to this result.
Truncating the expansion (\ref{c5b}) after the terms linear in $\om_L$, and inserting the result into Eq.(\ref{d4}) yields
\begin{eqnarray}\label{d9}
\T(\N,\psi_I) = \om_L Q(j)\Tau^2 (\N,\Om,\psi_I)+O(\om_L^2), 
\end{eqnarray}
with
\begin{eqnarray}\label{d99a}
 Q(j)\equiv \sum_{k=0}^{2j}\phi_k|\la \beta^k| \hat{j}_z|\beta_0\ra |^2.\q\q
\end{eqnarray}
In Eq.(\ref{d9}), the new time parameter $\Tau(\N,\Om,\psi_I)$, called the { \it SWP time} until a better name is found, 
is given by  \cite{FOOTswp}
\begin{eqnarray}\label{d10}
\Tau^2(\N,\Om,\psi_I) =W(\N,\psi_I)^{-1}\times \n
\sum_{N \in \N} W(N,\psi_I) |\ttau(N,\psi_I)|^2,\q\q
\end{eqnarray}
where 
\begin{eqnarray}\label{d10a}
W(N,\psi_I)\equiv |\la N|\hat{U}_{part}(t_2,t_1)|\psi_I\ra|^2
\end{eqnarray}
is the probability for the particle to be found in $|N\ra$ in 
the absence of the clock, and 
\begin{eqnarray}\label{d10aa}
W(\N,\psi_I)\equiv \sum_{N \in \N}W(N,\psi_I)
\end{eqnarray} 
is this probability for the whole of the chosen subspace $\N$.
\newline
Equation (\ref{d10}) is the central result of our discussion so far. In its l.h.s., we have an average, obtained for an ensemble weakly coupled SWP clocks, which is the observed result of the measurement. 
In the r.h.s., the only quantity describing the particle is  $\Tau(\N,\psi_I)$. It is given by the weighted sum of the squared moduli of the complex times defined in (\ref{b5}), evaluated for the transitions into all orthogonal states spanning the chosen subspace $\N$ of the particle's Hilbert space. This is an illustration of the general principle discussed in Sect. V: in a weakly perturbing inaccurate measurement the system is always represented by combinations of the relevant probability amplitudes, 
given in this case by $\overline{\tau}_{\Om}(N,\psi_F)$ of Eq.(\ref{b5}). 
\newline
We note also that the squares of the SWP times, rather than the SWP times themselves, are additive.
For two disjoint subspaces $\N$ and $\N'$, $\PP(\N)\PP(\N')=0$, equation (\ref{d4})  gives
\begin{eqnarray}\label{d11}
\Tau^2(\N\cup \N',\Om,\psi_I)=[W(\N,\psi_I)+W(\N',\psi_I)]^{-1}\times\q\q\n
[W(\N,\psi_I)\Tau^2(\N,\Om,\psi_I)+W(\N',\psi_I)\Tau^2(\N',\Om,\psi_I).
\end{eqnarray}
Next we  look at the SWP times from a slightly different prospective. 
\section{Calibration and the Uncertainty Principle}
A perhaps less direct way to arrive at the SWP time $\Tau(\N,\Om,\psi_I)$ is to calibrate SWP clock by using a procedure similar to the one proposed by Leavens in \cite{Leav1}. Consider first introducing a magnetic field everywhere in space, rather than just inside $\Om$.
Now all Feynman paths spend in the field the same duration $\tau=t_2-t_1$,
we have $A(\psi_F, \psi_I, t_2,t_1|\tau)=A(\psi_F, \psi_I, t_2,t_1) \delta(\tau-t_2+t_1)$ for all $|\psi_I\ra$ and $|\psi_F\ra$,
 and the clock decouples from the particle's motion. Thus, in the limit $\om_L\to 0$  this "free running" clock would measure
 [cf. Eq.(\ref{d9})] 
\begin{eqnarray}\label{dz1}
{\Tf}^{free}(t_2-t_1) = \om_L Q(j)(t_2-t_1)^2.
\end{eqnarray}
Next, let us measure  $\T(\N,\psi_I)$ in the case where the magnetic field exists only inside $\Om$. Let us also assume that there is some hypothetical
duration $\tau_{"in \Om"}$ which the particle spends in $\Om$, and during which the spin rotates. If so, the resulting value 
of $\T(\N,\psi_I)$ should be the same as ${\Tf}^{free}(\tau_{"in \Om"})$, obtained for a spin that has been in free rotation for $\tau_{"in \Om"}$ seconds.  Equating $\T(\N,\psi_I)$ in Eq.(\ref{d9}) to ${\Tf}^{free}(\tau_{"in \Om"})$ shows that we must  identify
the sought $\tau_{"in \Om"}$ with the SWP time,
\begin{eqnarray}\label{dz2}
\tau_{"in \Om"}=\Tau(\N,\Om,\psi_I). 
\end{eqnarray}
This may seem reasonable, since $\Tau(\N,\Om,\psi_I)$ is a real valued positive quantity, yet there is a serious concern.
Apparently, we {\it impose} a single duration while, according to Sect. V, there shouldn't be one.  It ought to be prudent to make further checks.
\newline
First we consider a classical case, where the magnetic field is confined to $\Om$, and the single path which connects $\psif$ with $\psii$, spends $\tau_{cl}$ seconds in $\Om$. Clearly,
$A(\psi_F, \psi_I, t_2,t_1|\tau)=A(\psi_F, \psi_I, t_2,t_1) \delta(\tau-\tau_{cl})$, and we obtain the correct result
\begin{eqnarray}\label{dz3}
\tau_{"in \Om"}=\tau_{cl}. 
\end{eqnarray}
The problems begin once quantum interference starts to play a role.
To see it, suppose that  there are exactly two virtual paths, via which $\psif$ can be reached from $\psii$ (perhaps in a situation similar to the one shown in Fig.1, or in a setup where the particle's wave packet is split into two parts, which pass via different optical fibres, and are later recombined). The paths spend in $\Om$ $\tau_1$ and $\tau_2$ seconds, and have the probability amplitudes $A_1$ and $A_2$, respectively. How much time does the particle spend in $\Om$?
Before proceeding, we recall the  Uncertainty Principle, already outlined in Sect. IV. Interference merges the two virtual paths into a single 
route connecting the particle's initial and final states. The duration spent in $\Om$ is, therefore, truly indeterminate \cite{FUNC2}. It is not $\tau_1$ or $\tau_2$, nor any other similar duration. It should not exist.
\newline
Noting that  now $A(\psi_F, \psi_I, t_2,t_1|\tau)=A_1 \delta(\tau-\tau_1)+A_2 \delta(\tau-\tau_2)$, [in the situation shown in Fig.1, the amplitude $A(x_F, x_I, t_2,t_1|\tau)$ has two stationary regions around $\tau_1$ and $\tau_2$, and is highly oscillatory elsewhere],
 and using (\ref{dz2}) we find
\begin{eqnarray}\label{dz4}
\tau_{"in \Om"}=
\left | \frac{A_1\tau_1+A_2\tau_2}{A_1+A_2}\right |
\end{eqnarray}
There are no {\it a priori} restrictions on the magnitude or the sigh of the ratio $A_2/A_1$. Suppose the transition takes three seconds, $t_2-t_1=3$, and the paths spend in $\Om$  $1$ and $2$ seconds, respectively. Choosing $A_1=0.5$ and 
$A_2= -A_1+0.001$ we find 
\begin{eqnarray}\label{i5}
\tau_{"in \Om"}=4998\text{s} >> t_2-t_1=3\text{s},
\end{eqnarray}
which is strange.
Further, choosing $A_1=0.5$ and $A_2=-0.25$, yields
\begin{eqnarray}\label{dz6}
\tau_{"in \Om"}=0,
\end{eqnarray}
which is also strange, especially if $\psii$ and $\psif$ are, as in Fig.1, localised on the opposite sides 
of the region $\Om$, which the particle, therefore, has to cross.
\section{Complex times and the "weak measurements"}
Both experiments described at the end of the previous Section,  can be performed, at least in principle. We must, therefore, decide on an interpretation of the results (\ref{i5}) and (\ref{dz6}).
There are two possibilities. Either, (A),  the $\tau_{"in \Om"}$ represents a physical duration, and has further implications
for our understanding of quantum motion.
Or, (B),  it is something else, in which case we need to explain what it is precisely. Next we look at both options.

A. The first option may appear either absurd \cite{Bohm} or intriguing, depending on the reader's viewpoint.
Indeed, should the $4998\text{s}$ in Eq.(\ref{i5}) be "physical", we must conclude that 
quantum mechanics allows a particle "to spend more than an hour in some place during a journey that last only three seconds".
By taking the $0\text{s}$ result in Eq.(\ref{i5}) literally, we defy Einstein's relativity by letting the particle "cross the region infinitely fast".
\newline
Both conclusions are reminiscent of other "surprising" results obtained within the so-called "weak measurements"
approach \cite{WEAK11}.
Among these one encounters the notions of "negative kinetic energy" \cite{NKE}, "negative number of particles" \cite{AhHARDY}, "having one particle in several places simultaneously" \cite{AhBOOK}, a "photon disembodied from its polarisation" \cite{CAT}, 
an "electron with disembodied charge and mass" \cite{CAT}, "an atom with the internal energy disembodied from the mass" \cite{CAT}, and "photons found in places they neither enter or leave" \cite{v2013}, \cite{vPRL}.
Perhaps closest to the subject of this paper is the concept of "charged particles moving faster than light through the vacuum", introduced in \cite{Cher}. In all these examples, the analysis relies on obtaining the "weak value"  $\la \B \ra_w$, of an operator $\B=\sum_i |b_i\ra B_i\la b_i|$, 
defined for a system prepared (pre-selected) in a state $\psii$ and then found (post-selected) in a state $\psif$, 
\begin{eqnarray}\label{wm1}
\la \B \ra_w=\frac{\la \psi_F|\B|\psi_I\ra}{\la \psi_F|\psi_I\ra}.
\end{eqnarray}
The conclusions of Refs. \cite{NKE}-\cite{Cher}, mentioned above, are then drawn from the properties of the complex valued quantity $\la \B \ra_w$.

B. Another explanation available to us is more down to earth (see also \cite{FUNC2}, \cite{PLA2016}, \cite{PLA2015},  \cite{PLAvaid}).
 The results (\ref{i5}) and (\ref{dz6}) may simply illustrate the Uncertainty Principle (above) by showing that it is impossible to ascribe 
a meaningful duration to a situation where two or more durations interfere to produce the result.
Response of a quantum system to an attempt to measure the traversal time without perturbing the system's motion results in evaluation 
of the sum of the probability amplitudes (\ref{b3}), which can, in principle, take any value at all. 
The weak value (\ref{wm1}) is another illustration of this general principle. Indeed, the final state $\psif$ can be reached
via passing through the eigenstates $|b_i\ra$, and the amplitude for the $i$-th virtual path is $A(i,\psi_f,\psi_i)=\la \psi_F|b_i\ra\la b_i|\psi_I\ra$.
Thus, Eq.(\ref{wm1}) can be cast in the form similar to (\ref{b5})
\begin{eqnarray}\label{wm1a}
\la \B \ra_w=\sum_{paths}B_iA(i,\psi_I,\psi_F)/\sum_{paths}A(i,\psi_I,\psi_F).
\end{eqnarray}
where the eigenvalues of $\B$ on the routes $\psif \gets |b_i\ra \gets \psii$, $B_i$, replace the values of $\tau_\Om[x(t)]$ on the virtual Feynman paths. 
We could call the complex time (\ref{b5})  "the weak value of the traversal time functional",
bearing in mind that it is {\it just} a particular combination of the probability amplitudes $A(\psi_F,\psi_I,t_2, t_1|\tau)$. 
No other interpretation of the complex times should be possible, as the Uncertainty Principle does not allow to view probability amplitudes, or their combinations, as the actual values of a physical quantity \cite{FUNC2}.
\newline
We strongly advocate the second explanation.  
Indeed, there is nothing strange about the result (\ref{dz4}) itself, and the only thing at fault is our desire to impose, through the calibration procedure,  a single physical duration $|\ttau|$ where there shouldn't be one. 
Not surprisingly, the result is unsatisfactory.
Accordingly, with Eq.(\ref{i5}), the clock has not "aged" by more than an hour in a span of three seconds. Its final state $|\gamma(t)\ra$ is a superposition of the states $|\gamma(t|\tau_1)\ra$ and $|\gamma(t|\tau_2)\ra$, rotated by the angles $\om_L \tau_1$ and
$\om_L \tau_2$, respectively. It is certainly not equal to $|\gamma(t|\tau_{"in \Om"})\ra$, and cannot, in general, be obtained by a rotation through  {\it any} angle $\omega_L \tau_{X}$,
\begin{eqnarray}\label{i6}
A_1|\gamma(t|\tau_1)\ra + A_2|\gamma(t|\tau_2)\ra \ne const  |\gamma(t|\tau_X)\ra,
\end{eqnarray}
since higher orders in $\om_L$ would involve quantities $\overline{\tau^n_\Om}$ in Eq.(\ref{xy5}), 
and $\overline{\tau^n_\Om}\ne (\ttau)^n$.
\newline
We must conclude that the calibration of a weak SWP clock fails to define a meaningful traversal time for a quantum particle. 
This is in agreement with the Uncertainty Principle.

\begin{figure}
	\centering
		\includegraphics[width=8.5cm,height=5.5cm]{{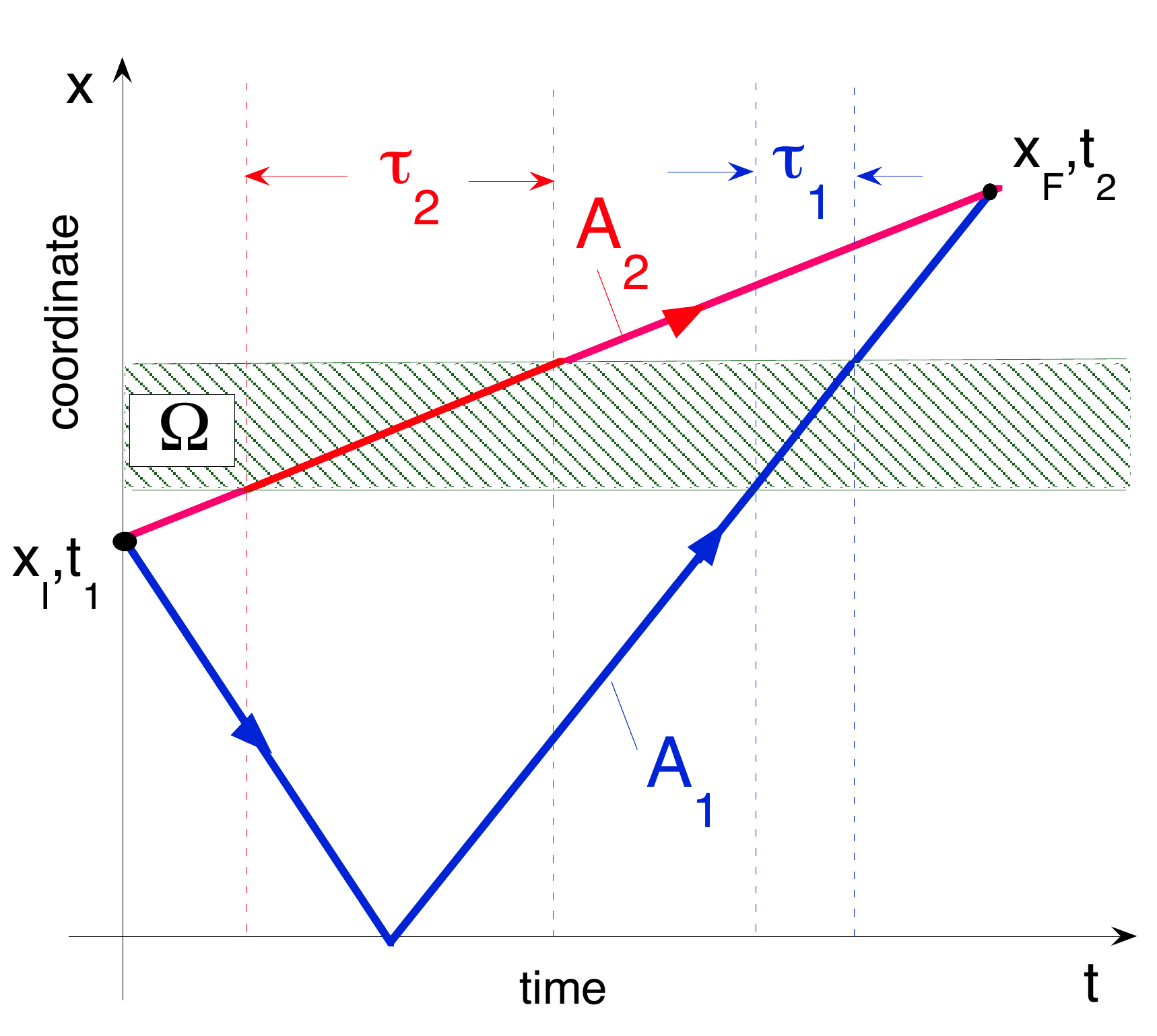}}
\caption{(Color online) A semiclassical particle can reach the final state $|x_F\ra$ from $|x_I\ra$ directly, and by having been reflected off a wall at $x=0$.  The particle is heavy, so there are two "classical" trajectories trajectories, which minimise the 
action $S$ in Eq.(\ref{b1}). The trajectories interfere, and
are travelled with the amplitudes $A_2$ and $A_1$, spending in the region $\Om$ $\tau_2$  $\tau_1$ and $\tau_2$ seconds, respectively. All in all, how much time does the particle spend in $\Om$? This is the "which way?" question at the centre of the traversal time controversy. } 
\label{fig:1}
\end{figure}
\section{The dwell time}
 Despite the difficulties outlined in the previous Section, much of the discussion about tunnelling times continues to be  built around a tacit assumption that a single classical-like duration, which characterises a classically forbidden transition, exists and simply has not yet been found \cite{LansREV}.  
 One candidate for the role of this parameter is the {\it dwell time}, a special case of the complex time (\ref{b5}), evaluated for the final state 
 $\psif$ obtained by unperturbed evolution of the initial state $\psii$,  
 \begin{eqnarray}\label{ga1a}
\tau^{dwell}_\Om(\psi_I) \equiv \ttau(\up(t_2,t_1)\psi_I,\psi_I).
\end{eqnarray}
 It can be written is several different ways. Expanding $|\up(t_2,t_1)\psi_I\ra$ is some basis $|N\ra$, we can express $\tau^{dwell}_\Om$ 
 in a form similar to Eq.(\ref{d10}),
 \begin{eqnarray}\label{ga2}
\tau^{dwell}_\Om(\psi_I)=\sum_{N} W(N,\psi_I) \ttau(N,\psi_I).\q\q
\end{eqnarray}
In the operator notations of Sect. VI, the dwell time becomes
 \begin{eqnarray}\label{ga2a}
\tau^{dwell}_\Om(\psi_I)=\la \psi_I|\up^\dagger(t_2,t_1)\up^{(1)}(t_2,t_1)|\psi_I\ra,
\end{eqnarray}
 and using Eq.(\ref{xy9}) yields a derived result, often mistaken for the definition of $\tau^{dwell}_\Om(\psi_I)$,
 \cite{Buett2, DSB,Leav1, Leav2}, 
\begin{eqnarray}\label{ga3}
\tau^{dwell}_\Om = \int_{t_1}^{t_2} dt'\int_{\Om}dx |\psi_I(x,t')|^2.
\end{eqnarray}
\newline
The dwell time possesses several attractive properties. 
Firstly, like its classical counterpart, it is non-negative, and never exceeds the total duration of motion, $\tau^{dwell}_\Om \le t_2-t_1$, thus avoiding the problems encountered in the previous Section.
Secondly, written as in Eq.(\ref{ga3}), it appears to have a simple probabilistic structure, with the contribution from the interval $dt'$  proportional
to the probability to find the particle in $\Om$. Thirdly, the same expression arises in approaches as different as the Bohm trajectories method \cite{Leav3}, and the Feynman path approach considered here. 

Before accepting $\tau^{dwell}_\Om$ as the long sought classical-like duration, we note 
that certain questions about it remain unanswered. The interference between different durations, which contribute to the transition, remains intact, and the conflict with Uncertainty Principle continues unresolved. Also, the said properties do not extend to the individual terms
$\ttau(N,\psi_I)$ in Eq.(\ref{ga2}), which remain complex-valued, and should not be confused with meaningful durations, as was shown in the previous Section. Finally, it may be that, for some unknown reason, a classical-like duration can only be defined for a quantum system, which
follows uninterrupted evolution along its "orbit" in the Hilbert space, $|\psi(t)\ra=\up(t,t_1)|\psi_i\ra$.
But then, to be a true analogue of the classical traversal time, $\tau^{dwell}_\Om(\psi_I)$ should arise whenever a time measurement perturbs the particle only slightly, and no post-selection is performed on the particle in the end. Should it not be so, the appealing form of Eq.(\ref{ga3}) would be fortuitous, and have no further physical consequences.  We will test this last assumption next.
\section{Would the SWP clock measure the dwell time?}
An example at hand is the SWP clock in the $\om_L \to 0$ regime. Suppose we run an ensemble of clocks between $t_1$ and $t_2$ 
without controlling the particle's final state, evaluate the average $\Tau(all,\psi_I)$ by choosing $\N$ to coincide with all of the particle's Hilbert space, and use the calibration procedure of Sect. X.
Will the result coincide with the dwell time (\ref{ga3}) as was assumed in \cite{SWP2}? 
The question was studied also in \cite{Leav1}.
\newline
The answer is yes, provided the system evolves along single classical path. We have already shown 
[see Eq.(\ref{dz3})] that
the SWP time coincides with the $\tau_{cl}$, evaluated for this path. With an (almost) classical particle represented by a 
very narrow wave packet crossing the region $\Om$, the equality $\tau^{dwell}_\Om=\tau_{cl}$ follows directly from the "stopwatch expression" (\ref{ga3}). Thus, $\tau_{cl}$ is the unique duration which arises from both approaches in the classical limit.
\newline 
However, in the full quantum case, the answer is no. From Eq.(\ref{d10}) we have 
\begin{eqnarray}\label{gc1}
\Tau(all,\Om,\psi_I) =
\la \psi_I|\up^{(1)\dagger}(t_2,t_1)|\up^{(1)}(t_2,t_1)|\psi_I\ra^{1/2},\q\q
\end{eqnarray}
and the application of (\ref{A10}) yields
\begin{eqnarray}\label{h3}
\Tau(all,\Om,\psi_I)
=\sqrt{\tau^{dwell}_\Om(\psi_I) \times \ttau(\psi^{(1)}(t_2),\psi_I)}\q
\end{eqnarray}
which, in general, is not the same as 
$\tau^{dwell}_\Om(\psi_I)$.
\newline
It is easy to see the reason for this discrepancy. According to Eq.(\ref{ga2a}) the dwell time must involve the product $\up^\dagger(t_2,t_1)\up^{(1)}(t_2,t_1)$,
and could only appear
in the linear in $\om_L$ corrections to  $P(k,\all)$ in Eq.(\ref{d5}). But with the choice of the states $|\beta^k\ra$ in Eq.(\ref{d2}), all such corrections vanish, 
for $k\ne 0$ because $\la \beta^k|\beta^0\ra=0$, and for $k=0$ since $\la \beta^0|\hat{j}_z|\beta^0\ra=0$.
Neither would  $\tau^{dwell}_\Om(\psi_I)$ appear  in the higher order corrections to $P(k,\all)$, since none of these corrections contain the required term $\up^\dagger(t_2,t_1)$.
 In general, the dwell time plays no role in the analysis of the SWP clock, as defined in Sect. VIII. However, in some special cases, $\Tau(all,\Om,\psi_I)$ may 
accidentally reduce to $\tau^{dwell}_\Om(\psi_I)$, as we will show in the next Section.
\section{Stationary tunnelling and the Leavens' analysis}
All that was said above applies to tunnelling of a particle prepared in a wave packet state, shown in the diagram in Fig.2.
The particle's initial state at $t=t_1$ is a superposition of the plane waves with positive momenta $p>0$,
\begin{eqnarray}\label{e1}
\psi_I(x)=\la x | \psi_I\ra=\int_0^\infty A(p)\exp(ipx) dp,
\end{eqnarray}
located to the left of a barrier of a finite width, $V(x)$, occupying the region $[0,d]$. The wave packet moves towards the barrier, and the energies $E(p)=p^2/2\mu$ are chosen all to lie below the barrier height, so that in order to be transmitted the particle has to tunnel.
At a sufficiently large time $t_2$, the scattering is complete, and the wave packet is divided into the transmitted (T), $\psi^T(x,t_2)$, 
and reflected (R), $\psi^R(x,t_2)$, parts,
\begin{eqnarray}\label{e2a}\nonumber
\la x |\psi(t_2)\ra\equiv\la x|\hat{U}_{part}(t_2-t_1)|\psi_I\ra=\q\q\q\q\q\q\n
\psi_T(x,t_2)+\psi_R(x,t_2)=\q\q\q\q\q\q\n
(2\pi)^{-1/2}\int_0^\infty dp T(p)A(p)\exp[px-iE(t_2-t_1)]+\q\q\\
(2\pi)^{-1/2}\int_0^\infty dp R(p)A(p)\exp[-ipx-iE(t_2-t_1)],\q\q
\end{eqnarray}
where $T(p)$ and $R(p)$ are the transmission and reflection amplitudes, respectively. 
\newline
We are interested in the duration spent in the barrier region, and choose $\Om\equiv [0,d]$.
In the limit $t_{1,2} \to \mp \infty$, matrix elements of the operators in Sect. VI between the plane waves $|p\ra$, 
$\la x|p\ra=exp(ipx)$, are given by ($n=0,1,2,...$)
\begin{eqnarray}\label{e3}
\la p'|\up^{(n)}(t_2-t_1)|p\ra =\q\q\q\q\q\q\q\q\q\n
 i^n\exp[-iE(p)(t_2-t_1)] \partial_\lambda^nT(p,\lambda=0) \delta(p-p'),\n
\q\text{for}\q p>0\q \text{and}\q p'>0,\n
\text{and}\nn
i^n\exp[-iE(p)(t_2-t_1)] \partial_\lambda^nR(p,\lambda=0) \delta(|p|-|p'|),\n 
\q\text{for}\q p>0\q \text{and}\q p'<0,
\end{eqnarray}
where $T(R)(p,\lambda)$ denote the transmission (reflection) amplitudes for a composite barrier $V(x)+\lambda\Theta_{[0,d]}(x)$.
From Eq.(\ref{xy6}), for the complex  tunnelling and reflection times of a particle with an initial momentum $p$ we have
\begin{eqnarray}\label{e4}
\tttu(p,p) = i\partial_\lambda \ln T(p,\lambda=0), 
\end{eqnarray}
and 
\begin{eqnarray}\label{e5}
\tttu(-p,p) = i\partial_\lambda \ln  R(p,\lambda=0). 
\end{eqnarray}
 In the following we will be interested only in whether the particle is transmitted, or reflected, and employ the projectors $\PP(\tunn)=\int_{0}^\infty |p\ra\la p|$ and $\PP(\refl)=\int_{-\infty}^0 |p\ra\la p|$, on all positive and all negative momenta, to distinguish between the two outcomes. 
\newline
Suppose next that a weak SWP clock is used to measure the time the particle spends in $\Om$. 
 To obtain the SWP time for transmission, 
we replace in Eq.(\ref{d10}) summation over $N$ by integration over $p$, and $\PP(\N)$ with $\PP(\tunn)$. We find
\begin{eqnarray}\label{e2}
\Tau(\tunn,[0,d],\Psi_I)=\nn
W(\tunn)^{-1/2}\left [\int_{0}^\infty dp |T(p)|^2 |A(p)|^2 |\tttu (p,p)|^2\right ]^{1/2},
\end{eqnarray}
and similarly for reflection, 
\begin{eqnarray}\label{e7}
\Tau(\refl,[0,d],\Psi_I)=\nn
W(\refl)^{-1/2}\left [ \int_{0}^\infty dp |R(p)|^2 |A(p)|^2 |\tttu(-p,p)|^2\right ]^{1/2},
\end{eqnarray}
where $W(\tunn)\equiv\int_{0}^\infty|T(p)|^2 |A(p)|^2 dp $ and $W(\refl)\equiv\int_{0}^\infty |R(p)|^2 |A(p)|^2 dp $, 
are the tunnelling and reflection probabilities, respectively. Finally, if we do not care whether the particle is transmitted or reflected, 
from Eqs.(\ref{d10}) and (\ref{e3}) we find the calibrated SWP result, without post-selection, to be
\begin{eqnarray}\label{e8}
\Tau(\all,[0,d],\Psi_I)=\nn
\left \{\int_{0}^\infty dp |A(p)|^2[|\partial_\lambda T(p,\lambda=0)|^2
 + |\partial_\lambda R(p,\lambda=0)|^2]\right \}^{1/2}
\end{eqnarray}
Returning to the question of the previous Section, we  may want to compare this with the dwell time which, according to Eqs.(\ref{ga2}), (\ref{e4}) and (\ref{e5}), is given by
\begin{eqnarray}\label{e9}
\tau^{dwell}_{[0,d]}(\psi_I)=\nn
i\int_{0}^\infty |A(p)|^2[T^*(p)\partial_\lambda T(p,\lambda=0)
 + R^*(p)\partial_\lambda R(p,\lambda=0)]dp
\end{eqnarray}
As expected, the SWP result in Eq.(\ref{e8}) is different from the dwell time in (\ref{e9}).
Leavens \cite{Leav1} studied, mostly numerically, the case where the incident particle has a definite momentum $p_0$.
To arrive at his results from Eqs.(\ref{e8}) and (\ref{e9}) it is sufficient to choose a nearly monochromatic wave packet, so narrow in the momentum space, that the transmission and reflection amplitudes and their derivatives can be approximated by their values at $p_0$.
Since $\int_0^\infty |A(p)|^2dp=1$, this yields 
\begin{eqnarray}\label{e10}
\Tau(\all,[0,d],p_0)=\nn
\left [|\partial_\lambda T(p_0,\lambda=0)|^2
 + |\partial_\lambda R(p_0,\lambda=0)|^2\right ]^{1/2}
\end{eqnarray}
and 
\begin{eqnarray}\label{e11}
\tau^{dwell}_{[0,d]}(p_0)=\nn
i[T^*(p_0)\partial_\lambda T(p_0,\lambda=0)
 + R^*(p_0)\partial_\lambda R(p_0,\lambda=0)]
\end{eqnarray} 
In \cite{Leav1} it was shown that a good agreement between the SWP result $\Tau(\all,[0,d],p_0)$ 
and  the dwell time (\ref{e11}) is achieved 
for free motion, if the width of the region, $d$, is sufficiently large. Good agreement between the two was also found for a barrier turned into a potential step,
e.g., if $d$ is sent to infinity, thus making transmission impossible.  
The latter result follows immediately by putting in Eqs.(\ref{e10}) and (\ref{e11}) $T(p_0,\lambda)\equiv 0$ and $|R(p_0)|\lambda\equiv 1$, which gives 
\begin{eqnarray}\label{e11a}
\Tau(\all,[0,\infty],p_0)=\nn
-\partial_\lambda Arg[R(p_0,\lambda=0)]=\tau^{dwell}_{[0,\infty]}(p_0).\q\q\q
\end{eqnarray} 
The case of a free particle crossing the region of a width $d$, requires a little more attention. We can neglect the reflection term in Eq.(\ref{e11}) but not in (\ref{e10}), and must evaluate both derivatives instead. Using Eq.(\ref{xy9}) we easily find
\begin{eqnarray}\label{i9}\nonumber
\frac{\Tau(\all,[0,d],p_0)}{\tau^{dwell}_{[0,d]}(p_0)}=\sqrt{1+\left | \frac{\sin(p_0d)}{p_0d}\right |^2}_{p_0d\to \infty}\to 1,
\end{eqnarray}
which explains the good agreement found by Leavens for broad regions.
Finally, another minor point regarding the analysis of \cite{Leav1} is consigned to the Appendix.
\newline
To summarise, we can agree with Leavens on the general discrepancy between the dwell time, and what is measured by a calibrated SWP clock. We also have dexplained why this discrepancy must arise. However, we disagree
with the final conclusion of \cite{Leav1} that "it is only the dwell time, which does not distinguish between transmitted and reflected particles, that is a meaningful concept in conventional interpretations of quantum mechanics". The dwell time is, we argue, just a special case of the "complex time" and is no more, and no less, meaningful than the tunnelling and reflection times in Eqs.(\ref{e4}) and (\ref{e5}).
\begin{figure}
	\centering
		\includegraphics[width=8.5cm,height=5.5cm]{{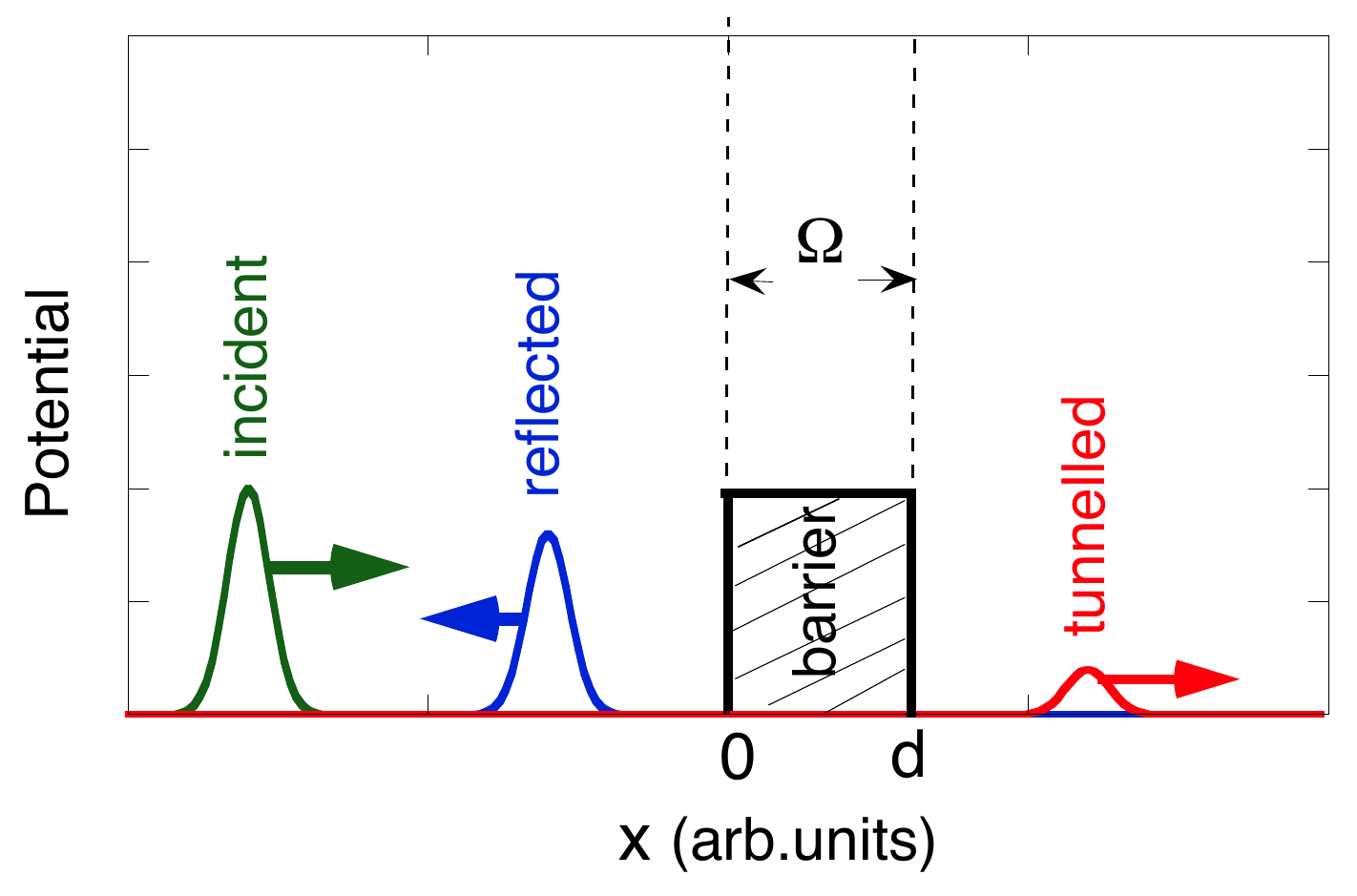}}
\caption{(Color online) An incident wave packet impacts on a potential barrier, and is divided into the transmitted (tunnelled) and reflected parts. What is the duration the particle has spent in the region $\Om$ which contains the barrier? } 
\label{fig:2}
\end{figure}
\section{Tunnel ionisation}
Our general analysis applies also to the case of tunnel ionisation, where the tunnelling time problem has attracted recent theoretical interest \cite{LansREV}.
In an ionisation experiment, an initially bound electron has a chance to escape by tunnelling across a potential barrier briefly created by a time dependent external field. One may be interested in the duration the escaped electron has spent in the classically forbidden region, and attempt to measure it by means of a weak SWP clock, perturbing tunnelling as little as possible. 
A realistic calculation of such a measurement can be found, for example, in \cite{SWP2}, and here we will limit ourselves to the formulation of the problem, and the identification of the time parameters such a measurement would produce. 
\newline
A one-dimensional sketch of the setup is shown in Fig.3. Bound at  $t=t_1$ in the single bound state of a potential well, $|\psi_I\ra=|\psi_{0}\ra$,  the particle can escape  to the continuum while an external field is converting the binding potential into a potential barrier. Long after the field is switched off, at some $t=t_2$, its wave function is divided into the "bound" part, describing the particles which failed to leave the well, and the "free" part,  describing the escaped particles moving away from it. We, therefore, have
\begin{eqnarray}\label{ti1}
\up(t_2,t_1)|\psi_I\ra=|\psi_{bound}\ra+|\psi_{free}\ra\equiv \n
C(t_2,t_1)]|\psi_{0}\ra+
\int_0^\infty B(p,t_2,t_1)|p\ra dp. 
\end{eqnarray}
and find the ionisation probability to be given by, 
\begin{eqnarray}\label{iprob}
W(\ion)=\la \psi_{free} | \psi_{free}\ra =\int_0^\infty |B(p,t_2,t_1)|^2dp.
\end{eqnarray} 
Let the particle be monitored by a weak SWP clock, with the magnetic field is localised in the classically forbidden region $\Om$, as shown in Fig.3.
We will also have at our disposal a perfect remote detector, capable of determining whether the particle has escaped, and if it has, and able to evaluate its momentum $p$. With this, we can choose to post-select the particle in the free state, and record the clock's reading only if the particle was seen to escape. We can also post-select it in the bound state, and keep the readings only in the case the remote detector has not fired. Alternatively, we can choose not to post-select at all, and retain all of the clock's readings.
\newline
There is a set of complex times which, as discussed in Sect. V, are related to the response of the system 
to the introduction of a constant potential $\La \Theta_\Om(x)$ in the region of interest.  If such a potential is introduced, 
the wave function at $t_2$ retains the form (\ref{ti1}), but its coefficients should depend on $\lambda$, $C(t_2,t_1)\to C(\La,t_2,t_1)$, $B(p,t_2,t_1)\to B(p,\La,,t_2,t_1)$.
Thus, for an escaped particle with a momentum $p$ we can define the complex time (\ref{b5}) and other complex "averages" (\ref{xy5}) as  [we omit the time dependence of the coefficients $C$ and $B$, and recall that $ \overline{\tau_\Om}\equiv  \overline{\tau^1_\Om}$]
\begin{eqnarray}\label{ti2}
 \overline{\tau^n_\Om}(p,\psi_0,t_2,t_1)=(i)^nB(p, \La=0)^{-1}\partial^n_\La B(p,\La=0).\q
\end{eqnarray}
Similarly, for a particle which remained in the well, we have
\begin{eqnarray}\label{ti2a}
 \overline{\tau^n_\Om}(\psi_0,\psi_0,t_2,t_1)=(i)^nC(\La=0)^{-1}\partial^n_\La C(\La=0).
\end{eqnarray}
There is also a real valued dwell time, which does not distinguish between 
the particles which have escaped and those which remained bound,
\begin{eqnarray}\label{ti3}
{\tau^{dwell}_\Om}(\psi_0)=
|C(\La=0)|^2  \overline{\tau_\Om}(\psi_0,\psi_0,t_2,t_1)\n 
+\int_0^\infty |B(p,\La=0)|^2 \overline{\tau_\Om}(p,\psi_0,t_2,t_1)dp.
\end{eqnarray}
\newline
What is measured in an  experiment depends on how the clock is prepared and read. 
If the weak SWP clock of Sect. IX is used, and the calibration procedure of Sect. X is applied, 
the time found for the particles which remain bound in the potential well is  
\begin{eqnarray}\label{ti4}
\Tau(\bound,\Om,\psi_0)=|\overline{\tau_\Om}(\psi_0,\psi_0,t_2,t_1)|\n
=|\partial_\La \ln C(\La=0)|.\q\q\q\q\q\q\q\q
\end{eqnarray}
For the particles which leave the well with unspecified momentum, the measurement will yield
\begin{eqnarray}\label{ti5}
\Tau(\free,\Om,\psi_0)=\q\q\q\q\q\q\n
W(\ion)^{-1/2}\left [{\int_0^\infty|\overline{\tau_\Om}(p,\psi_0,t_2,t_1)|^2 |B(p)|^2dp}\right ]^{1/2}\n
=W(\ion)^{-1/2}\left [{\int_0^\infty |\partial_\La B(p, \La=0)|^2dp}\right ]^{1/2}
\end{eqnarray}
Finally, if the final state of the particles is not controlled, from (\ref{d11}) we have
\begin{eqnarray}\label{ti6a}
\Tau(\all,\Om,\psi_0)=\{[1- W(\ion)]\times\q\q\q\q\n
\Tau^2(\bound,\Om,\psi_0)
+W(\ion)  \Tau^2(\free,\Om,\psi_0)\}^{1/2}\n
=\left [|\partial_\La C(\La=0)|^2+ \int_0^\infty |\partial_\La B(p, \La=0)|^2dp   \right ]^{1/2},
\end{eqnarray}
which is not the same as the dwell time in Eq.(\ref{ti3}).
\newline
If, on the other hand, we follow Leavens \cite{Leav1} in choosing $|\gamma^I\ra=|\beta^j\ra$,  (see Appendix),
the sum in the r.h.s. of Eq.(\ref{d99a}) will vanish, and to evaluate the new SWP time $\Tau'$, we would need to go to the next order 
in $\om_L$ in Eq.(\ref{c5b}).
For example, instead of (\ref{ti5}) from Eq.(\ref{Ap2}),  we will have
\begin{eqnarray}\label{ti4a}
\Tau'(\free,\Om,\psi_0) =\nn
\end{eqnarray}
\begin{eqnarray}\label{ti4aa}\nonumber
W(\ion)^{-1/3}
 \left \{ {\int_0^\infty \text{Re}[ \ttau(p,\psi_0)\overline{\tau^2_\Om}^*(p,\psi_0)] |B(p)|^2dp}\right \}^{1/3}=\n
W(\ion)^{-1/3} \left \{ {\int_0^\infty \text{Im}[\partial_\La B(p,\La=0) \partial_\La^2 B^*(p,\La=0)dp}\right \}^{1/3}.
\end{eqnarray}
and, as before,  will not recover the dwell time (\ref{ti3}) in the case no post-selection is made. 
\newline
The dwell time would, however, occur naturally if instead of evaluating the averages (\ref{d4}) or (\ref{Ap2}), we would 
employ a more general Larmor clock, described in Sect. VII, and consider a small difference in the probability
$P(k,\all)\equiv \la \Psi(t_2)|\beta^k\ra\la \beta^k|\Psi(t_2)\ra$ for the clock to be found in a state $|\beta^k\ra$
before and after it interacts with the particle. A simple calculation, using Eq.(\ref{c5b}), shows that this change 
is proportional to $\tau^{dwell}_\Om(\psi_0)$
\begin{eqnarray}\label{ti5a}
\delta P(k,\all) \equiv P(k,\all)-|\la \beta^k|\gamma^I\ra|^2=\n
2\om_L  \text{Im}[\la \gamma^I|\beta^k\ra\la \beta^k|\hat{j}_z|\gamma^I\ra] \tau^{dwell}_\Om(\psi_0)+O(\om_L^2)
\end{eqnarray}
Defining the measured mean value as $\delta \T(\all,\psi_I)\equiv \sum_{k=0}^{2j}\tau_k \delta P(k,\all)$, 
we obtain
\begin{eqnarray}\label{ti6}
\delta \T(\all,\psi_I)= Q'(j) \tau^{dwell}_\Om(\psi_0)+O(\om_L), 
\end{eqnarray}
with $Q'(j)= 2\text{Im}\left \{\sum_{k=0}^{2j} \phi_k\la \gamma^I|\beta^k\ra\la \beta^k|\hat{j}_z|\gamma^I\ra \right \}$.
If the magnetic field is introduced everywhere in space, we find $\delta \T^{free}(t_2-t_1)=Q'(j)(t_2-t_1)$.
Using this relation to calibrate the result (\ref{ti6}), as was done in Sect. X, shows that for this particular clock, 
the duration imposed in the quantum case is $\tau^{dwell}_\Om(\psi_0)$.
This is the case of linear calibration, studied by Leavens and McKinnon in \cite{Leav2}.
\newline
Thus, also in the case of tunnel ionisation, application of  a weak SWP clock does not yield a single real duration
the particle is supposed to spend in the classically forbidden region, 
but rather a variety of complex valued time parameters, through which the real valued result of the measurement is expressed.
These parameters differ for different settings of the clock,
and reduce to a unique classical value only in the primitive semiclassical limit, 
where a single classical trajectory connects the initial and final states.
In the next Section we give our conclusions.
\begin{figure}
	\centering
		\includegraphics[width=8.5cm,height=5.5cm]{{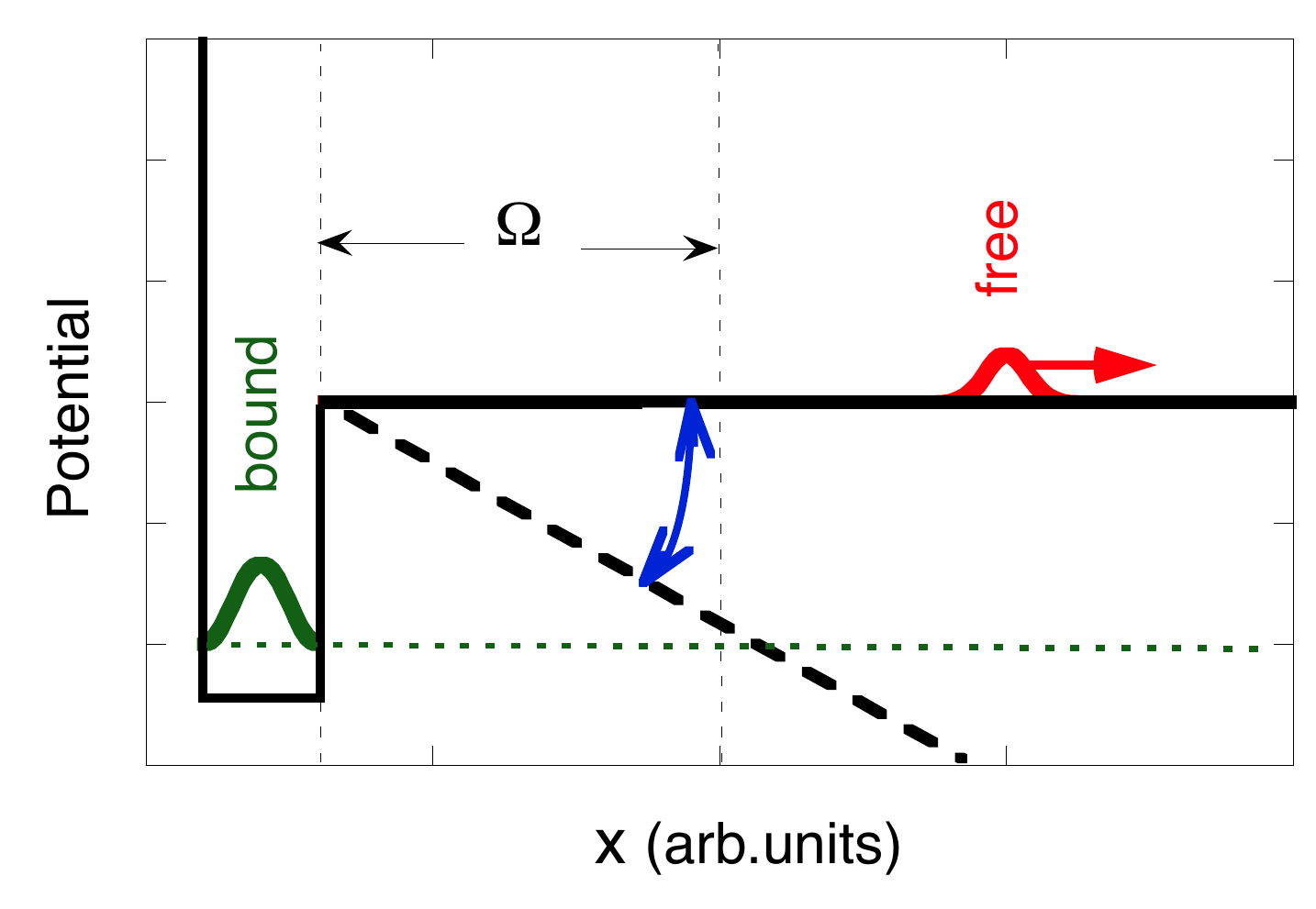}}
\caption{(Color online) at first the particle is trapped in the ground state of a potential well.
A time dependent external field turns the potential step into a barrier, and then restores it to its original shape.
The particle's wave function is divided into the part still trapped, and the escaped part, freely propagating away from the well.
What is the duration the particle has spent in the classically forbidden region $\Om$? } 
\label{fig:3}
\end{figure}
\section{Conclusion and discussion}
Mathematical exercises presented above do not themselves form a basis for a discussion about "the amount of time a tunnelling particle spends in the barrier".
They only illustrate the far more general principle at stake. Most of the quantum transitions, and certainly tunnelling, 
are interference phenomena, which require contributions from many virtual Feynman paths.
Each Feynman path spends certain amount of time, $\tau_\Omega[path]$, inside the region of interest $\Om$. We can group together the paths 
sharing the same value of  $\tau_\Omega[path]$, and see a transition as a result of interference between all traversal times involved.
The difficulty in determining the duration, spent by a quantum particle in $\Om$, is then the well known difficulty in answering the 
"which way?" ("which $\tau$?) question in the presence of interference, the only mystery in quantum mechanics, according to Feynman \cite{FeynL}. In this paper we have examined in detail one particular way of trying to answer the question, while leaving the interference intact. Arguably, the general conclusions, which can be drawn from our analysis, are more important  than any of its technical details. We will formulate these conclusions in a perhaps unusual form of attempting to ask the most relevant questions, and then trying to answer them the best we can.  

{\it a) What is measured by the SWP clock?} Like every clock of the Larmor family, the SWP clock measures the net time $\tau$ the particle's Feynman paths spend in the region of interest. 

{\it b) How is this time measured?} By modifying the contributions of different $\tau$'s to the particle's transition amplitude, 
depending on final state in which the clock is observed.

{\it c) Does the SWP analysis come up with an "operator for the tunnelling time"?}
Strictly speaking, no. The operator (\ref{d4a}), often quoted in that capacity, acts on the variables of the clock, 
and not on the variables of the particle. It defines, therefore, a von Neumann measurement which needs 
to me made on the spin. 

{\it d) To what accuracy is it measured?} If the function $G_{SWP}$ in Eq.(\ref{c4}) limits the values of $\tau$, which contribute to the 
transition $\psii|\beta^0\ra \to \psif|\beta^k\ra$, to a region of a width $\Delta \tau$ around some value $\tau_k$, we can say that
by observing the clock in $|\beta^k\ra$, we have measured a value $\tau_k$ to an accuracy $\Delta \tau$. 
A weak ($\om_L\to \infty$) SWP clock, whose main purpose is to perturb the transition as little as possible, 
does not discriminate between different times in this way. Rather, it studies the response of a particle to the small variations 
of the probability amplitudes defined in Eq.(\ref{b3}), and its accuracy is very poor. 

{\it e) Is there a probability distribution for the traversal time in the case of tunnelling?}
Not unless it is created by an accurate clock, which destroys the interference between different values of $\tau$.
If a weak clock is employed, only the {\it probability amplitude distribution} $A(\psi_F, \psi_I, t_2,t_1|\tau)$  in Eq.(\ref{b3})
is available.

{\it f) Are complex traversal times inevitable?} Interfering (virtual) pathways should together be considered a single route connecting the  initial and 
final states of the system. By the Uncertainty Principle \cite{FeynL}, virtual pathways cannot be distinguished without destroying 
interference between them. Accordingly, the response of a system to a weakly perturbing measurement of the traversal time functional (\ref{a1})
is always formulated in terms 
of the complex valued sum of the corresponding amplitudes, $A(\psi_F, \psi_I, t_2,t_1|\tau)$ in Eq.(\ref{b3}), weighted by the values of the functional, $\tau$ \cite{FUNC2}.
This is a general result behind the so-called weak measurement theory \cite{WEAK11}, \cite{PLA2016}. The complex time $\ttau$ in (\ref{b5}) is the "weak value" of the functional (\ref{a1}). 

{\it g) Can complex times be measured?} Certainly, for example by a weak SWP clock discussed above, and the fact that they are  complex valued is no major obstacle.
However, since the result of a measurement must be real, it is impossible to say {\it apriori} whether a particular experiment 
would yield $\text{Re} \ttau$, $\text{Im} \ttau$, $|\ttau|$, or, indeed, any other real valued combination of $\text{Re} \ttau$ and $\text{Im} \ttau$
(see also \cite{DSwp}). Our detailed analysis of the SWP clock used by Peres \cite{SWP3}, shows that what it measures is, in fact, $|\ttau|$.

{\it h) Are complex times related to physical time intervals?} In general, they are not. Any attempt at over-interpretation, by treating parts of $\ttau$ as if they were actual durations, would lead to insurmountable difficulties. For example, in the case described in Sect. X, one would face not only the chance of faster-than-light travel, but also the possibility of spending a month on the beach during a one-week leave from office. None of the two are offered by elementary quantum mechanics.

{\it i) What are the complex times then?} Just what their definition tells us. A complex time is what one would obtain by multiplying the amplitude to reach $\psif$ from $\psii$ and spend a duration $\tau$ in $\Om$ by $\tau$, and sum over all the $\tau$'s which contribute to the transition. 

{\it j) Is the dwell time more meaningful than other complex times?} No, it is a particular case of a complex time, whose 
attractive properties can be traced to the fact that the operator $\up(t_2,t_1|\lambda)$ in Eq.(\ref{xy3}) is hermitian for any real $\lambda$ \cite{NEGAT}. In the quantum case, it does not always take the place of the classical duration, as was shown in Sect. XII.
The Uncertainty Principle does not forbid the weak values to look appealing in particular cases. Rather, it guarantees the existence
of "unappealing" results, should different initial and final states be chosen instead \cite{PLA2015}. It is these other results which should warn one against giving too much credit to the nice exceptions.

{\it k) Does the SWP clock measure the dwell time?} 
As defined in \cite{SWP3} and in Sect. VIII, it does not. The choice of the states in which the clock is observed is such that the terms which add up to
 the dwell time do not contribute to the result, even if the final state of the particle is not controlled. 
With a different Larmor clock it would, however, be possible to evaluate $\tau^{dwell}_\Om(\psi_I)$ \cite{Leav2}.

{\it l) Does tunnelling particle spend a finite amount of time in the barrier?} We could equally ask whether the electron in the Young's double slit experiment reaches the screen 
by passing through the holes in the screen? All Feynman paths which contribute to tunnelling spend some time in the barrier.
Moreover, replacing the Schroedinger equation with a relativistic Klein-Gordon one \cite{Low}, leaves only the paths 
spending in $\Om$ a time longer than the $\frac{width\q of\q the\q region}{speed\q of \q light}$ \cite{DSrel}. In every {\it virtual} scenario (i.e., the one to which we can ascribe an amplitude, but not the probability \cite{FUNC2} ) the electron goes through one of the holes, and the particle spends a reasonable duration inside the barrier.

{\it m) How much time does a tunnelling particle spend in the barrier?} We could equally ask "which hole did the electron go through?".
In standard (Feynman) quantum mechanics it goes through both, and through neither one in particular \cite{FeynL}. 
In the same sense, the particle spends in the barrier all durations at the same time.
The question is meaningless in a very strong sense, and an attempt to force it brings an unsatisfactory answer, $\overline{\tau}_{\Om}(\psi_I,\psi_F)$.
Consider two researchers using two weak Larmor clocks, but one determining $\text{Re}\ttau$, and the other $|\ttau|$, 
for a transition where $\text{Re}\ttau$ is zero, but $|\ttau|$ is not. 
To the first researcher the transition takes no time in $\Om$, to the second researcher this time is finite.
Their subsequent argument would have no resolution, as both would be right about their results, 
but both will be wrong in their final conclusions. 

{\it n) Can one expect the complex time (\ref{b5})  to occur in other applications? } Only where the quantity of interest
can be obtained by integrating the amplitudes $A(\psi_F, \psi_I, t_2,t_1|\tau)$ over $\tau$. Some examples were given in 
\cite{DSwp} and \cite{SBrouard}.

{\it o)  Can there be other definitions of the tunnelling time?} In quantum mechanics, the failure to define one unique tunnelling time
does not mean that such times cannot be defined at all. On the contrary, it means that there are more possible time parameters, than in the classical case \cite{STEIN}. Firstly, there are $\text{Re} \ttau$, $\text{Im} \ttau$, $|\ttau|$ already mentioned. Then there are weak values of other functionals, e.g., of $\Ttu[\x(t)]$ in Eq.(\ref{a2}). There are also times not related to Feynman paths. One famous example is the phase time \cite{REV5}, which can be interpreted as the weak value of the spacial shift with which the particle leaves the scatterer, divided by the particle's velocity \cite{DSann}. Moreover, one can define other times, e.g., as the moments the front, the maximum, the rear, or the centre of mass of a wave packet passes through a chosen surface in space \cite{REV1}. The Pollack and Miller time \cite{PM}, and the times mentioned in Sect. III,  provide further examples.

{\it p) Can there be a unique tunnelling time scale?} That is, could one leave aside all the details of the previous discussion, and simply be assured that tunnelling takes approximately  $\tau_{approx}$ microseconds, so that all devices using it should not go faster that $\tau_{approx}$? The answer in standard (Feynman) quantum mechanics appears to be "no". If there were such a time scale, it could be found
by examining
 the corresponding amplitude distribution $A(\psi_F, \psi_I, t_2,t_1|\tau)$. 
For example, for a particle of a given energy,
 tunnelling across a rectangular barrier, the amplitude distribution 
 is oscillatory, and exhibits a fractal behaviour \cite{SBrouard}. Hence, its Fourier spectrum contains all frequencies, and we cannot associate with it any specific time scale {\it a priori}. In a particular application, $A(\psi_F, \psi_I, t_2,t_1|\tau)$ may be
integrated with a smooth function $G(\tau)$, whose width $\Delta \tau$  determines  which of the higher frequencies would be neglected. However, the process of making $\Delta \tau$ smaller will never converge to a result which no longer depends on $\Delta \tau$. Thus, we argue, any new tunnelling time measured in an experiment, or found theoretically, should be used strictly in the particular context it was obtained. For instance, a statement "the peak of the tunnelled wave packet has arrived at the detector $1$ fs. earlier than that of a free propagating one" is correct. Its extension "... and, therefore, the particle has spent $1$ fs. less in the barrier" is unwarranted.  Any claim to find the universal tunnelling time, or time scale, is likely to be misleading.

{\it q) And the classical time scale?} One exception to $p)$ is the (semi) classical limit, where rapidly oscillating $A(\psi_F, \psi_I, t_2,t_1|\tau)$ develops a very narrow stationary region around single classical value  $\tau_{cl}$ \cite{SBrouard}. If so, the contribution to any (within reason) integral over $\tau$, involving $A(\psi_F, \psi_I, t_2,t_1|\tau)$, comes from the vicinity of $\tau_{cl}$. Appearance of a single stationary region signals, therefore, return to the classical description.

{\it r) Could an extension, or alternative formulation of quantum mechanics help define the traversal time in a different way?} 
Such a theory will have also solved the "which way?" problem for the double-slit experiment.

{\it s) Did Bohm's trajectories approach achieve that?.} 
One approach which claims to achieve that is the Bohm' causal interpretation \cite{Holl}, \cite{Bohm2}.
In Bohm's theory, a particle moves along a streamline of a probability current calculated with a time dependent wave function 
$\psi(x,t)$, and its initial position is distributed according to $|\psi(x,t=0)|^2$. The streamlines cannot cross, 
and a Bohm's trajectory leading to a given point on the screen in the Young's experiment always passes through one of the slits. 
Similarly, a particle crossing a region of space always spends there a unique amount of time.
A detailed comparison between the Bohm's  trajectory and the Feynman path approaches to the tunnelling time problem was made in \cite{Leav3}, where the author concluded that the two approaches are incompatible. 
It is not our purpose to continue this discussion, and we will limit ourselves to just two comments.
Firstly, the unperturbed Bohm's trajectories do not help us with the analysis of the SWP clock, while the Feynman paths do.
Bohm's trajectories are formulated in the absence of a measuring device, and must change once such a device is introduced,
in order to describe its effects.
Secondly, by using Feynman amplitudes, one can define the time any quantum system spends in an arbitrary subspace of its Hilbert space. For example we can define and measure the time a qubit spends in one of its states \cite{DSresid}, \cite{DSprl2}.
It is unclear how Bohm's approach can be extended to cover these cases. 

In summary, we have analysed the work of a weakly perturbing  Salecker-Wigner-Peres clock in terms of virtual Feynman paths, and related it to the complex traversal time first introduced in \cite{DSB}. We have shown that in the standard (Feynman) quantum mechanics the appearance of complex times in an inevitable consequence of the Uncertainty Principle. We also explained why these complex times, or their real valued combinations, should not be interpreted as physical durations, and tried to draw some of more general conclusions about the state of the tunnelling problem in quantum theory. 

\section{Appendix: A different choice of the initial state for an SWP clock}
It is worth clarifying one difference between our results of Sect. X  and those of \cite{Leav1}.
 According to Eq.(26) of  \cite{Leav1}, for a free running clock,
as $\om_L\to 0$, we must have  $\T^{free}\sim \om_L^2$,  whereas according to our Eq. (\ref{dz1}) is should be proportional to 
$\om_L$. 
The reason is that in \cite{Leav1} Leavens considered also choosing a different initial state for the clock, replacing
($j$ is an integer)  $|\beta^0\ra$ with $|\beta^j\ra$, and effectively postulated a {\it negative} duration $\tau'_{k-j}=(\phi_k-\phi_j)/\om_L<0$ each time the clock is found in $|\beta^k\ra$ with $0\le k <j$. In this case, from Eq.(\ref{d2a}) we have
$G_{SWP}(\om_L\tau|j,\beta^k,\beta^j)=G_{SWP}(\om_L\tau|j,\beta^{k-j},\beta^0)$, and  $|\beta^{k-j}\ra \equiv\exp[-i\hat{j}_z(\phi_k-\phi_j)]|\beta^0\ra$, so that Eq.(\ref{d4}) becomes 
\begin{eqnarray}\label{Ap1}
\T'(\N,\psi_I)=\sum_{k=0}^{2j}\tau'_{k-j} P(k-j,\N), 
\end{eqnarray}
which is also Eq.(20) of \cite{Leav1}. Proceeding as in Sect. IX, we find that, with this choice, the contribution to $\T'(\N,\psi_I)$, linear in $\om_L$, vanishes, leaving 
$\T'(\N,\psi_I)$ proportional to $\om_L^2$ as $\om_L\to 0$. For a freely running clock, with the magnetic field introduced everywhere in space, we have $\T'^{free}(t_2-t_1)\sim (t_2-t_1)^3$. Calculating $\T'(\N,\psi_I)$ to the first non-vanishing order in $\om_L$, and comparing the result with $\T'^{free}(t_2-t_1)$, we find that the time $\Tau'(\N,\Om,\psi_I)$, measured by the modified clock, is given by
\begin{eqnarray}\label{Ap2}
\Tau'(\N,\Om,\psi_I) =W(\N,\psi_I)^{-1/3}\times \q\q\q\q\n
\left \{ \sum_{N \in \N} W(N,\psi_I) \text{Re}[ \ttau(N,\psi_I)\overline{\tau^2_\Om}^*(N,\psi_I)] \right \}^{1/3},\q\q
\end{eqnarray}
which involves also the complex valued square of the functional (\ref{a1}), defined in Eq.(\ref{xy5}).
\newline

 \section {Acknowledgements}  Support of
MINECO and the European Regional Development Fund FEDER, through the grant
FIS2015-67161-P (MINECO/FEDER) 
is gratefully acknowledged.

 

\begin{thebibliography}{10}
\bibitem{FOOTbox} To simplify the notations we  assume that the particle is in a large box, so its states $|N\ra$ are discrete.
 As the size of the box tends to infinity, the sums over $N$ should be replaced by the corresponding integrals.
\bibitem{FOOTswp} Note that $W(N,\psi_I)=0$ does not guarantee that the corresponding term
in the sum in Eq.(\ref{d10}) vanishes, since $|\ttau(N,\psi_I)|^2=\la \psi^{(1)}|\psi^{(1)}\ra/W(N,\psi_I)$.
\end{thebibliography}
\end{document}